\documentclass[journal=jacsat,manuscript=article]{achemso}
\usepackage{amsmath,amssymb}%equations and symbols
\usepackage{tikz}%plot figure
\usepackage{braket}%Dirac bracket
\usepackage{algorithm,algorithmic}
\usepackage{multirow}%table multirow
\usepackage{enumerate}
\usepackage{graphicx}
\usepackage{threeparttable}
\usepackage{natbib}
\usepackage{soul,color,xcolor}

\SectionNumbersOn
\author{Shizhe Jiao}
\affiliation{Hefei National Research Center for Physical Sciences at the Microscale, and Anhui Center for Applied Mathematics, University of Science and Technology of China, Hefei, Anhui 230026, China}
\author{Jielan Li}
\affiliation{Hefei National Research Center for Physical Sciences at the Microscale, and Anhui Center for Applied Mathematics, University of Science and Technology of China, Hefei, Anhui 230026, China}
\email{jielanli@mail.ustc.edu.cn}
\author{Xinming Qin}
\affiliation{Hefei National Research Center for Physical Sciences at the Microscale, and Anhui Center for Applied Mathematics, University of Science and Technology of China, Hefei, Anhui 230026, China}
\author{Lingyun Wan}
\affiliation{Hefei National Research Center for Physical Sciences at the Microscale, and Anhui Center for Applied Mathematics, University of Science and Technology of China, Hefei, Anhui 230026, China}
\author{Wei Hu}
\affiliation{Hefei National Research Center for Physical Sciences at the Microscale, and Anhui Center for Applied Mathematics, University of Science and Technology of China, Hefei, Anhui 230026, China}
\email{whuustc@ustc.edu.cn}
\author{Jinlong Yang}
\affiliation{Key Laboratory of Precision and Intelligent Chemistry, and Department of Chemical Physics, University of Science and Technology of China, Hefei, Anhui 230026, China}
\title[CPX]
  {
  Complex-valued K-means clustering of interpolative separable density fitting algorithm for large-scale hybrid functional enabled \textit{ab initio} molecular dynamics simulations within plane waves
  }%\footnote{Supporting Information Available}}

\begin{document}

\makeatletter
\setlength\acs@tocentry@height{6.5cm}
\setlength\acs@tocentry@width{6.1cm}
\makeatother

%\begin{tocentry}	
%\begin{center}
%    \includegraphics[width=\textwidth]{Fig/toc.pdf}
%\end{center}
%\end{tocentry}
 
\begin{abstract}

K-means clustering, as a classic unsupervised machine learning algorithm, is the key step to select the interpolation sampling points in interpolative separable density fitting (ISDF) decomposition. 
Real-valued K-means clustering for accelerating the ISDF decomposition has been demonstrated for large-scale hybrid functional enabled \textit{ab initio} molecular dynamics (hybrid AIMD) simulations within plane-wave basis sets where the Kohn-Sham orbitals are real-valued. 
However, it is unclear whether such K-means clustering works for complex-valued Kohn-Sham orbitals. 
Here, we apply the K-means clustering into hybrid AIMD simulations for complex-valued Kohn-Sham orbitals and use an improved weight function defined as the sum of the square modulus of complex-valued Kohn-Sham orbitals in K-means clustering. 
Numerical results demonstrate that this improved weight function in K-means clustering algorithm yields smoother and more delocalized interpolation sampling points, resulting in smoother energy potential, smaller energy drift and longer time steps for hybrid AIMD simulations compared to the previous weight function used in the real-valued K-means algorithm. 
In particular, we find that this improved algorithm can obtain more accurate oxygen-oxygen radial distribution functions in liquid water molecules and more accurate power spectrum in crystal silicon dioxide compared to the previous K-means algorithm. 
Finally, we describe a massively parallel implementation of this ISDF decomposition to accelerate large-scale complex-valued hybrid AIMD simulations containing thousands of atoms (2,744 atoms), which can scale up to 5,504 CPU cores on modern supercomputers.

\end{abstract}

\section{Introduction}\label{sec:Introduction}

The transposed Khatri-Rao product~\cite{slyusar1998end, Sankhya_30_167_1968} (also known as face-splitting product) $Z =\{z_{ij}:= \phi_i(\mathbf{r})\psi_j^\ast(\mathbf{r})\}_{1 \leq i \leq N_{\phi}, 1 \leq j \leq N_{\psi}} \in \mathbb{C}^{N_r \times (N_{\phi}N_{\psi})}$ of Kohn-Sham orbitals $\phi_i(\mathbf{r})$ and $\psi_j(\mathbf{r})$ in real space $\{\mathbf{r_i}\}_{i = 1}^{N_r}$ is inevitable for the multi-center integrals of advanced electronic structure calculations in density functional theory (DFT),~\cite{hohenberg1964inhomogeneous,kohn1965self} especially for the Hartree-Fock (HF)~\cite{slater1951simplification,becke1993new} and post-HF electronic structure theory, such as time-dependent density functional theory (TDDFT),~\cite{Casida1995,PhysRev_96_951_1954} GW approximation~\cite{hedin1965new, hybertsen1986electron, aryasetiawan1998gw, onida2002electronic} plus Bethe-Salpeter equation (BSE),~\cite{salpeter1951relativistic} Second-order Møller-Plesset perturbation theory (MP2),~\cite{haser1993moller,feyereisen1993use,bernholdt1996large} and random phase approximation (RPA).~\cite{ren2012random} In order to reduce such high computational cost and memory usage of such multi-center integrals in the Kohn-Sham DFT calculations, several low rank approximation algorithms have been proposed, such as the Cholesky decomposition,~\cite{Beebe_1977,Roeggen_1986} resolution of identity (RI),~\cite{Whitten_1973,Dunlap_1979,Vahtras_1993,Weigend_2002} tensor hypercontraction (THC)~\cite{THC_2012,Parrish_2012,Hohenstein_2012} and pseudospectral decomposition.~\cite{Friesner_1985,Friesner_1987} However, it is difficult to apply these low rank approximation algorithms to the cases of atomic forces and vibrational frequencies especially for periodic systems within plane-wave basis sets, which can be used for a wide range of applications such as geometry optimization and \textit{ab initio} molecular dynamics (AIMD) simulation. 
% The memory usage of representing such transposed Khatri-Rao product on grids points $\{\mathbf{r}_{i}\}_{i=1}^{N_{r}}$ in real space is $O(N_rN_{\phi}N_{\psi})$.
%These low-rank approximation algorithms can efficient compress the transposed Khatri-Rao product of Kohn-Sham orbitals

%And the computational cost of compressing the transposed Khatri-Rao product is $O(N_r^2N_{\phi}N_{\psi})$ for most of these low-rank approximation algorithms.
% The orbital pairs set is more complex as the investigated system becomes large, which limits seriously the speed of the DFT calculations. 
Recently, Lu et al. proposed a new tensor hypercontraction (THC) algorithm by using the randomized QR factorization with column pivoting (QRCP) procedure~\cite{lu2015compression}, namely interpolative separable density fitting (ISDF),~\cite{lu2017cubic, hu2017interpolative} which can achieve an effectively low rank approximation of the transposed Khatri-Rao product of the Kohn-Sham orbitals ($\phi_i$ and $\psi_j$) and compress their redundant information with cubic-scaling computational cost of $O(N_rN_{\phi}N_{\psi})$. The transposed Khatri-Rao product of the Kohn-Sham orbitals can be expressed by $z_{ij}= \phi_i(\mathbf{r})\psi_j^\ast(\mathbf{r}) \approx \sum_{\mu = 1}^{N_{\mu}}\zeta_{\mu}(\mathbf{r})\phi_i(\mathbf{r_{\mu}})\psi_j^\ast(\mathbf{r_{\mu}})$, where $\{\mathbf{r}_{\mu}\}_{\mu=1}^{N_{\mu}}$ are a set of interpolation points from grid points $\{\mathbf{r}_{i}\}_{i=1}^{N_{r}}$ in real space, $N_{\mu}$ is proportional to $\sqrt{N_{\phi}N_{\psi}}$ ($N_{\mu} = t \sqrt{N_{\phi}N_{\psi}}$, $t$ is the rank truncation constant), and $\zeta_{\mu}(\mathbf{r})$ is the auxiliary basis functions (ABFs). The ISDF decomposition has already been applied successfully in several types of multi-center integrals in the Kohn-Sham DFT calculations within Gaussian-type orbitals (GTOs), numerical atomic orbitals (NAOs), and plane-wave (PW) basis sets, such as hybrid DFT calculations,~\cite{hu2017interpolative,dong2018interpolative,qin2020interpolative,qin2020machine} RPA correlation,~\cite{lu2017cubic} quantum Monte Carlo (QMC) simulations,~\cite{malone2018overcoming} TDDFT,~\cite{hu2020accelerating} MP2,~\cite{lee2019systematically} GW~\cite{gao2020accelerating,ma2021realizing} and BSE~\cite{hu2018accelerating} calculations, for molecular and periodic systems.
%In the past several decades, much efforts have been devoted to solving the Kohn-Sham (KS) equations to obtain wave function of many-body system,~\cite{hohenberg1964inhomogeneous,kohn1965self} which determines all ground state properties of the system in principle. Selection of exchange correlation functional has a decisive influence on the precision of solutions to Kohn-Sham equations. As is well-known, Jacob’s ladder~\cite{perdew2001jacob} is composed of five levels, including the local density approximation (LDA), the generalized gradient approximation (GGA), meta-GGA, hybrid exchange−correlation functionals and the random phase approximation (RPA), whose accuracy and complexity for electronic structure calculations increases in turn. Hybrid exchange-correlation functionals is required in many computing cases, such as calculations of band gap of semiconductors. It is necessary to solve $O(N_e^2)$($N_e$ is the number of electrons) Poisson-like equations when we use iterative method for KS equations with hybrid functionals, which is a big challenge for large scale systems including hundreds of atoms or even more. In order to reduce computational complexity of hybrid functionals, Interpolative separable density fitting (ISDF) can be introduced to lessen the number of Poisson-like equations to be solved from $O(N_e^2)$ to $O(N_e)$.~\cite{hu2017interpolative} ISDF~\cite{lu2015compression} is a fast and effective method to compress the redundant
%%approximation and so on. 

The ISDF decomposition can be divided into two key steps,~\cite{hu2017interpolative} including selecting the interpolation points (IPs) and computing interpolation vectors (IVs). The IVs can be computed easily by a least-squares fitting procedure when the IPs are selected. The IPs mean the selection of a set of nonuniform grid points $\{\mathbf{r}_{\mu}\}_{\mu=1}^{N_{\mu}}$ where the values of the orbital pairs evaluated are almost consistent with that evaluated in all grid points $\{\mathbf{r}_{i}\}_{i=1}^{N_{r}}$. Two approaches have been proposed to select the IPs. The standard approach is the randomized QRCP as mentioned previously,~\cite{lu2015compression} which is accurate but expensive in the ISDF decomposition process.~\cite{hu2017interpolative} Another approach is the centroidal Voronoi tessellation (CVT) algorithm proposed by Dong et al.,~\cite{dong2018interpolative} which only requires the information from the electron density in the DFT calculations. The CVT method can be performed easily by K-means clustering algorithm, a classical unsupervised machine learning algorithm. Such K-means clustering algorithm aims at dividing $N_r$ data points into $N_{\mu}$ clusters. Each cluster includes all points whose distances from the centroid of the cluster are smaller than that of other clusters. Since K-means clustering only converges to a local optimal solution, the accuracy of K-means clustering strongly depends on the selection of initial centroids and the definition of distance and centroids.~\cite{vassilvitskii2006k, singh2013k, dong2018interpolative, qin2020machine} For real-valued orbitals, recent numerical results~\cite{hu2017interpolative,dong2018interpolative,qin2020interpolative,qin2020machine} have demonstrated that K-means clustering algorithms with relatively simple weight definitions can yield reasonably high accuracy at a much lower cost. 

However, the Kohn-Sham orbitals are usually represented as complex-valued Bloch functions with k-point sampling~\cite{bloch1929quantenmechanik, bloch1957generalized, wu2021ace} and noncollinear spin density functional theory~\cite{chen2023low} for periodic systems and time-dependent wavefunctions~\cite{runge1984density, yabana1996time, onida2002electronic, andreani2005measurement, rttddftes} for real-time time-dependent density functional theory (RT-TDDFT). For example, the transposed Khatri-Rao product of complex-valued Kohn-Sham orbitals $Z =\{z_{ij}:= \sum_{k,l=1}^{N_g} \phi_{i,k}\psi_{j,l}^\ast e^{i(G_k-G_l) \cdot \mathbf{r}}\}$, where $N_g$ is the number of plane-wave basis sets. $\phi_{i,k}$ and $\psi_{j,l}^\ast$ are expansion coefficients within plane-wave basis sets for $\phi_i$ and $\psi_j^\ast$, respectively. In this case, conventional K-means clustering algorithms get into trouble~\cite{wu2021ace, chen2023low, rttddftes} because the centroids and distance cannot be well-defined for complex-valued points. 
An alternative solution is to separate the complex data into real and imaginary parts, then perform twice K-means clustering calculations and merge the real and imaginary centroids together. This solution does not work for complex-valued Kohn-Sham orbitals, since it is difficult to merge the real and imaginary parts for their one-to-one correlation. To the best of our knowledge, there are few works on complex-valued K-means clustering algorithms yet.~\cite{wu2021operational} In particular, Zhang et al. used complex-valued K-means to learn filters in conventional neural networks for sleep stage classification.~\cite{zhang2018complex} They defined complex-valued centroids and inner product instead of Euclidean distance. However, the complex number was converted from the real number in this work and the real and imaginary parts are dependent on each other. The approach is too expensive to be suitable for complex-valued Kohn-Sham orbitals in the DFT calculations to deal with the transposed Khatri-Rao product directly.
%Rahman et al. proposed a new quantum clustering method, which takes the magnitude square of a mixture wavefunction as the mixture distribution.~\cite{2016quantum} The method demonstrates more accurate results in an examples application of color segmentation.
%Compared with conventional K-means clustering algorithm, the K-means in ISDF decomposition process requires different computational methods to obtain the centroid of clusters, which is the weighted average of all points belong to the cluster because of the unequal weighted points in real space.

In the ISDF method, we introduce a weighted K-means clustering algorithm. Because the real and imaginary parts of complex-valued Kohn-Sham orbitals are discrete on the same grids in real space, we can perform K-means clustering on the grids in real space and define different centroids from that in conventional K-means clustering algorithm, which is defined as the weighted average of all points belonging to the cluster. We convert the information of complex-valued Kohn-Sham orbitals into the weight function. Therefore, it is important to choose an appropriate weight function to compute the weighted average for complex-valued Kohn-Sham orbitals. 

In this work, we present an improved weight function desirable for complex-valued Kohn-Sham orbitals in the DFT calculations. We apply successfully the improved K-means clustering algorithm into complex-valued hybrid functional calculations with plane-wave basis sets and achieve the acceleration of large-scale hybrid density functional calculations containing thousands of atoms. In particular, the \textit{ab initio} molecular dynamics for complex-valued hybrid DFT calculations in molecular and solid systems can be performed, such as liquid water molecules~\cite{santra2007accuracy,santra2007accuracy,santra2009coupled,santra2011hydrogen,santra2013accuracy,distasio2014individual,ko2020enabling,dong2018interpolative} and aluminium-silicon alloy,~\cite{khoo2011ab,zhang2017adaptive} which is important but expensive within plane basis sets.~\cite{chawla1998exact,mandal2021achieving, aimdhf0, aimdhf1, aimdhf2, aimdhf3} We demonstrate that the energy potential calculated using this improved weight function shows smoother, smaller energy drift and longer time steps compared to the previous weight function in the K-means algorithm. Therefore, this improved K-means clustering algorithm can accurately and efficiently accelerate large-scale and long-time AIMD with complex-valued hybrid DFT calculations. 

This work is organized as follows. Section~\ref{sec:Methodology} gives a brief description of the theoretical methodology, including the ISDF method, the complex-valued K-means clustering algorithm in ISDF, the combination of hybrid DFT calculations with ISDF method as well as their parallel implementation. Section~\ref{sec:Results} validates the numerical accuracy and computational efficiency of the complex-valued K-means clustering algorithm for ISDF decomposition to accelerate the hybrid DFT calculations. A summary and outlook is given in Section~\ref{sec:Conclusion}.

\section{Methodology} \label{sec:Methodology}

\subsection{Interpolative separable density fitting}\label{sec:ISDF}
%For the two-electron integrals, interpolative separable density fitting (ISDF) decomposition is an effective way to reduce computational cost. 
The ISDF decomposition is a new THC algorithm proposed by Lu and Ying firstly~\cite{lu2015compression} and then promoted by Hu et al. for large-scale hybrid DFT calculations~\cite{hu2017interpolative}. This algorithm can achieve low rank approximation of transposed Khatri-Rao product (also known as face-splitting
product) $Z =\{z_{ij}:= \phi_i(\mathbf{r})\psi_j^\ast(\mathbf{r})\}_{1 \leq i \leq N_{\phi}, 1 \leq j \leq N_{\psi}} \in \mathbb{C}^{N_r \times (N_{\phi}N_{\psi})}$ of Kohn-Sham orbitals $\phi_i$ and $\psi_j$.
\begin{equation}
\phi_i(\mathbf{r})\psi_j^\ast(\mathbf{r}) \approx \sum_{\mu = 1}^{N_{\mu}}\zeta_{\mu}(\mathbf{r})C_{\mu}^{ij}
\label{eq:decomposition}
\end{equation}
where the transposed Khatri-Rao product can be approximately decomposed into auxiliary basis functions (ABFs) $\zeta_{\mu}(\mathbf{r})$ and expansion coefficients $C_{\mu}^{ij}$. The number of auxiliary basis functions $N_{\mu}=t\sqrt{N_{\phi}N_{\psi}}$ can be regarded as the numerical rank of the decomposition, where $t$ is a small parameter to achieve the compromise between the numerical accuracy and computational efficiency. The key to ISDF decomposition is to solve the expansion coefficients. The tensor hypercontraction (THC)~\cite{hohenstein2012tensor,parrish2012tensor,Hohenstein_2012,parrish2013discrete} algorithms have made a success on it. The ISDF algorithm provides a new tensor hypercontraction to obtain the coefficients
\begin{equation}
C_{\mu}^{ij} = \phi_i(\mathbf{r_{\mu}})\psi_j^\ast(\mathbf{r_{\mu}})
\label{eq:Cmu}
\end{equation}
where $\{r_{\mu}\}_{\mu=1}^{N_{\mu}}$ are a set of interpolation points from grid points $\{r_{i}\}_{i=1}^{N_{r}}$ in real space. Therefore, the transposed Khatri-Rao product can be expressed in the following form
\begin{equation}
\phi_i(\mathbf{r})\psi_j^\ast(\mathbf{r}) \approx \sum_{\mu = 1}^{N_{\mu}}\zeta_{\mu}(\mathbf{r})\phi_i(\mathbf{r_{\mu}})\psi_j^\ast(\mathbf{r_{\mu}})
\label{eq:ISDFdecomposition}
\end{equation}

Thus the ISDF method can be divided into two steps.~\cite{hu2017interpolative} The first one is to get the expansion coefficients $C_{\mu}^{ij}$, namely to compute the IPs, which can be achieved by using the QRCP procedure or the weighted K-means clustering algorithm. From the matrix point of view, IPs procedure is to select $N_\mu$ rows from $Z$ for fitting the entire matrix $Z$. Because Z matrix is not full rank, the QRCP procedure can achieve the low rank decomposition of $Z$ as follows,
\begin{equation}
Z^T\Pi = Q\begin{bmatrix} R_{11} & R_{12} \\ 0 & 0 \end{bmatrix}_{N_{\phi}N_{\psi} \times N_r}
\label{eq:QRCP}
\end{equation}
where $Z^T$ is the transpose of matrix $Z$, $\Pi$ is the permutation matrix whose first $N_\mu$ columns give the IPs $\{\mathbf{r}_{\mu}\}_{\mu=1}^{N_{\mu}}$. $Q$ and $R_{11}$ denote orthogonal matrix and non-singular upper triangular matrix, respectively. The absolute values of the diagonal entries for matrix $R_{11}$ follow a non-increasing order. As the standard method, the computational cost of the QRCP procedure scales as $O(N_{\mu}^2N_r)$, while the cost of K-means clustering algorithm is $O(N_rN_{\mu})$, as Table~\ref{table:Cost} shows.

%\WH{Add the formulas of QRCP for IPs}

The second step is to get the ABFs $\zeta_{\mu}(\mathbf{r})$, namely to compute the IVs, which can be obtained by the least-squares procedure. After we obtain the expansion coefficients $C_{\mu}^{ij}$, eq \ref{eq:decomposition} can be written in matrix form
\begin{equation}
Z \approx \Theta C
\label{eq:decomMatrix}
\end{equation}
where $Z$ is $N_r \times N_e^2$ matrix ($N_{\phi}  \approx  N_{\psi}  \sim O(N_{e})$, $N_e$ is the number of electrons), which comes from $\phi_i(\mathbf{r})\psi_j^\ast(\mathbf{r})$ sampled on a set of dense real space grids $\{\mathbf{r}_{i}\}_{i=1}^{N_{r}}$. $\Theta = [\zeta_1, \zeta_2,..., \zeta_{N_{\mu}}]$, $N_r \times N_{\mu}$ matrix, namely IVs. $C = [\phi_i(\mathbf{r}_1)\psi_j^\ast(\mathbf{r}_1),...,\phi_i(\mathbf{r}_{\mu})\psi_j^\ast(\mathbf{r}_{\mu}),...,\phi_i(\mathbf{r}_{N_{\mu}})\psi_j^\ast(\mathbf{r}_{N_{\mu}})]^T$. Thus the IVs $\Theta$ can be given by
\begin{equation}
\Theta = ZC^T(CC^T)^{-1}
\label{eq:theta}
\end{equation}
where $ZC^T$ and $CC^{T}$ both need $O(N_e^4)$ floating point operations. Nevertheless, the separable structure of $Z$ and $C$ can reduce the operations dramatically.~\cite{hu2017interpolative} As is well known, 
\begin{equation}
\sum_{i,j}\phi_i\psi_j = (\sum_i \phi_i)(\sum_j \psi_j)
\label{eq:phipsi}
\end{equation}
Thus the $\mu$-th row, $\nu$-th column element of $ZC^T$ can be written as
\begin{equation}
%e_{\mu}^TZC^Te_{\nu} = 
P^{\phi}(\mathbf{r}_{\mu},\mathbf{r}_{\nu})P^{\psi}(\mathbf{r}_{\mu},\mathbf{r}_{\nu})
\label{eq:muRownuCol}
\end{equation}
where $P^{\phi}(\mathbf{r}_{\mu},\mathbf{r}_{\nu})$ and $P^{\psi}(\mathbf{r}_{\mu},\mathbf{r}_{\nu})$ are defined as 
\begin{equation}
  \begin{split}
P^{\phi}(\mathbf{r}_{\mu},\mathbf{r}_{\nu}) &= \sum_{i}^{N_{\phi}}\phi_i(\mathbf{r}_{\mu})\phi_i^\ast(\mathbf{r}_{\nu})  \\
P^{\psi}(\mathbf{r}_{\mu},\mathbf{r}_{\nu}) &= \sum_{i}^{N_{\psi}}\psi_i(\mathbf{r}_{\mu})\psi_i^\ast(\mathbf{r}_{\nu})
  \end{split}
\label{eq:muRownuColden}
\end{equation}
Here to compute $P^{\phi}$ and $P^{\psi}$ takes $O(N_e^3)$ floating point operations and the multiplication of $P^{\phi}$ and $P^{\psi}$ only needs $O(N_e^2)$ floating point operations. This conclusion also applies to $CC^T$. Therefore, we can reduce the computational complexity of IVs from $O(N_e^4)$ to $O(N_e^3)$.

\begin{table}[!htb]
\centering 
%\setlength{\tabcolsep}{0.5mm}
%\footnotesize
\caption{Computational cost and memory usage of IPs and IVs in ISDF decomposition. Notice that $N_r \approx 1,000 \times N_{e}$,
and $N_{\phi}  \approx  N_{\psi} \approx  N_{\mu} \sim O(N_{e})$ in the plane-wave
basis sets.} 
\begin{tabular}{cccc} \ \\
\hline \hline
Step & Algorithm  & Computation   &  Memory  \\
\hline
\multirow{2}*{IPs} &  QRCP& $O(N_{\mu}^2N_r)$ &  $O(N_rN_{\mu})$ \\
~ & K-means & $O(N_rN_{\mu})$  & $O(N_rN_{\mu})$  \ \\
\hline
IVs & Least-squares & $O(N_rN_{\mu}N_e)$ & $O(N_rN_{\phi}N_{\psi})$ \ \\
\hline \hline
\end{tabular} \label{table:Cost}
\end{table}

\subsection{Complex-valued K-means clustering in ISDF}\label{sec:KMeansforISDF}

As an unsupervised machine learning algorithm, the K-means clustering algorithm has been demonstrated to be much cheaper than the QRCP procedure.~\cite{dong2018interpolative, qin2020machine} Ideally, conventional K-means clustering algorithm is for seeking the solution to the following optimization problem.
\begin{equation}
arg min  \sum_{\mu=1}^{N_{\mu}}\sum_{\mathbf{r_k} \in C_{\mu}} ||Z(\mathbf{r}_k) - Z(\mathbf{r}_{\mu})||^2
\label{eq:opt}
\end{equation}
Here we divide the $N_r$ data points into $N_{\mu}$ clusters $\{C_{\mu}\}_{\mu=1}^{N_{\mu}}$. Each cluster can be denoted by its centroid, namely the IPs we need.
\begin{equation}
{C_{\mu}} = \{\ Z(\mathbf{r}_i) \ |\ dist(Z(\mathbf{r}_i), Z(\mathbf{r}_{\mu})) \leq dist(Z(\mathbf{r}_i), Z(\mathbf{r}_m))\ for\ all\ m \neq \mu\ \}
\label{eq:C_k}
\end{equation}
where the $dist(Z(\mathbf{r}_i), Z(\mathbf{r}_{\mu}))$ means distance between data points $Z(\mathbf{r}_i)$ and centroid $Z(\mathbf{r}_{\mu})$.
The optimization problem can only be solved by iterative calculations and K-means clustering converges to a local minimum. The accuracy of K-means clustering strongly depends on the selection of initial centroids and definition of distance and centroids.~\cite{vassilvitskii2006k, singh2013k, dong2018interpolative, qin2020machine} The conventional K-means clustering is mainly used for real-valued data points. As mentioned above, due to the dependence of the real and imaginary parts, there is a dilemma when we apply the conventional K-means clustering algorithm to the complex-valued Kohn-Sham orbitals. In addition, it is quite expensive to perform direct K-means clustering on the transposed Khatri-Rao product.

In the ISDF method, we effectively avoid these problems. We perform K-means clustering on the grids in real space where the real and imaginary parts of complex-valued Kohn-Sham orbitals map the same grids instead of the transposed Khatri-Rao product. APPENDIX A verifies the feasibility of this strategy. In ISDF method, the optimization problem is reduced to
\begin{equation}
arg min  \sum_{\mu=1}^{N_{\mu}}\sum_{\mathbf{r_k} \in C_{\mu}} w(\mathbf{r}_k)||\mathbf{r}_k - \mathbf{r}_{\mu}||^2
\label{eq:opt}
\end{equation}
where the $w(\mathbf{r}_k)$ is the weight function. The distances between the grid points are calculated by the Euclidean metric. The centroids $\mathbf{r}_{\mu}$ can be defined by the weighted average of all points that belong to the corresponding cluster as follows
\begin{equation}
\mathbf{r}_{\mu} = \frac{\sum_{\mathbf{r}_j \in C_{\mu}}\mathbf{r}_j w(\mathbf{r}_j)}{\sum_{\mathbf{r}_j \in C_{\mu}}w(\mathbf{r}_j)}
\label{eq:r_k}
\end{equation}
The information of complex-valued Kohn-Sham orbitals is converted into the weight function. Therefore, it is very important to select and define the empirical weight function. Different empirical weight functions have been proposed for different data types in practice.~\cite{dong2018interpolative, qin2020machine} 

For hybrid functional electronic structure calculations within numerical atomic orbitals (NAOs), we have proposed the norm of the row of orbital pairs as the weight function,~\cite{qin2020machine} namely
\begin{equation}
w(\mathbf{r}) = \sum_{i,j=1}^{N_b}|\varphi_i(\mathbf{r})||\varphi_j(\mathbf{r})| 
\label{eq:NAOs}
\end{equation}
%{\WH {cite papers and give some examples here}}
where $\{\varphi_i\}_{i=1}^{N_b}$ denote the real-valued NAOs and $N_b$ denotes the number of NAOs. It should be noticed that such two sets of real-valued orbitals involved in the transposed Khatri-Rao product are the same. 

For real-valued hybrid DFT calculations within plane-wave basis sets,\cite{dong2018interpolative} we have defined the weight function as
\begin{equation}
w(\mathbf{r}) = \sum_{i,j=1}^{N_e}|\phi_i(\mathbf{r})|^2|\psi_j(\mathbf{r})|^2 = (\sum_{i=1}^{N_e}|\phi_i|^2)(\sum_{j=1}^{N}|\psi_j|^2)
\label{eq:weight function}
\end{equation}
The weight function is the product of the square modulus (abbreviated as PSM) of real-valued Kohn-Sham orbitals $\{\phi_i\}_{i=1}^{N_e}$ and $\{\psi_j\}_{j=1}^{N}$, where $N$ is the number of Kohn-Sham orbitals. It should be noticed that such two sets of real-valued Kohn-Sham orbitals involved in the transposed Khatri-Rao product are different because a two-level self-consistent field (SCF) iteration procedure~\cite{hu2017adaptively} is used in hybrid DFT calculations within plane-wave basis sets.

\begin{figure}[!htb]
\begin{center}
\includegraphics[width=0.5\textwidth]{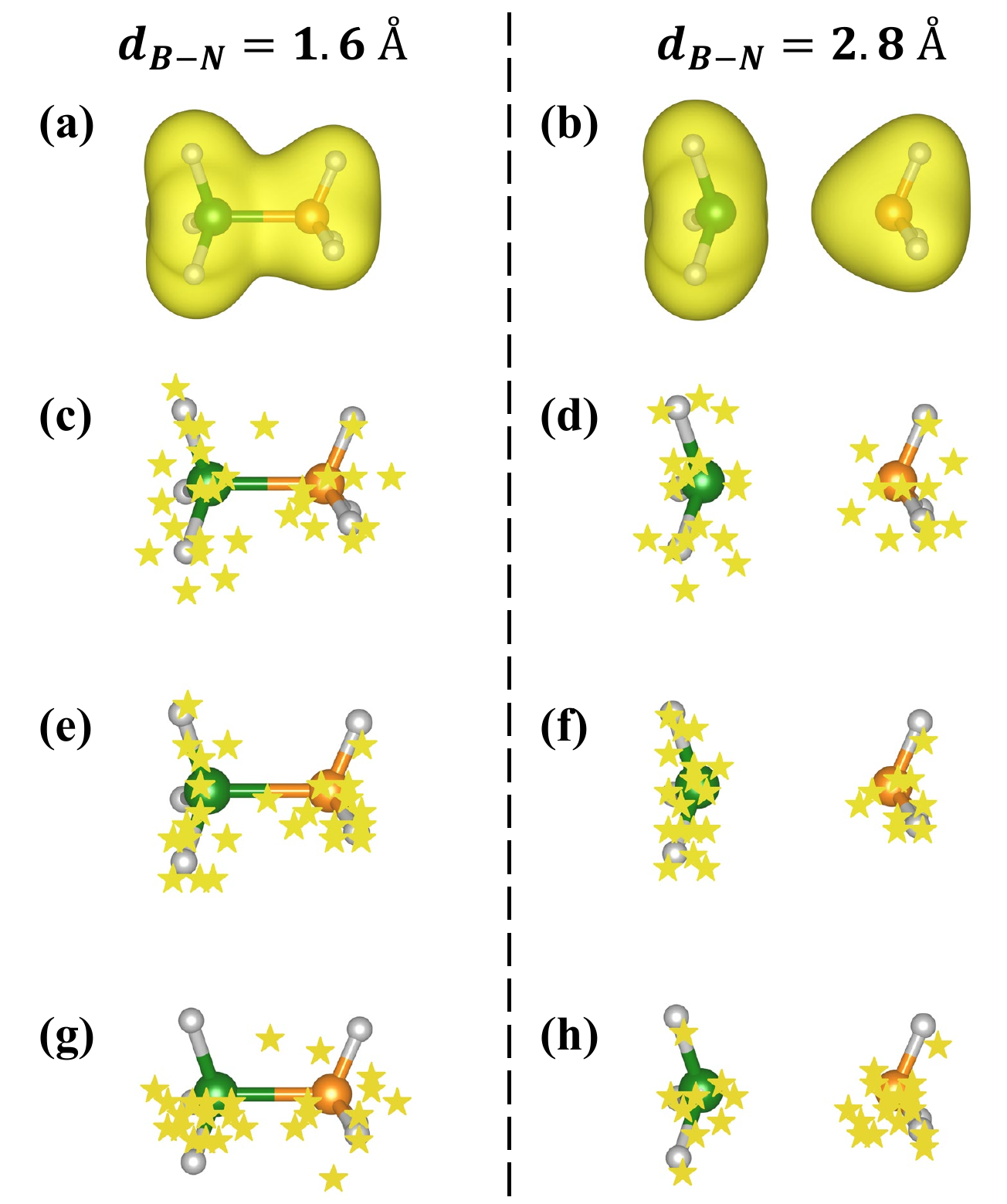}
\end{center}
\caption{Comparison of the IPs selected by the K-means with SSM and PSM as well as QRCP procedures for BH$_3$NH$_3$ with different B-N distances of $d_\textrm{B-N}$ = 1.6 \AA and 2.8 \AA, including (a, b) the electron density (yellow isosurfaces), (c, d) the IPs (yellow pentagrams) by the K-means with SSM, (e, f) the IPs (yellow pentagrams) by the K-means with PSM and (g, h) the IPs (yellow pentagrams) by the QRCP procedure.} \label{fig:ponits}
\end{figure}

However, the PSM weight function defined in Eq.\ref{eq:weight function} is prone to more zero elements, which makes the sampling points more localized in the real system as shown in FIG.~\ref{fig:ponits}1 (e, f), and the selected points are concentrated around the ball and stick model. Because the rows of Z selected at the IPs are linearly independent, the IPs should be delocalized as much as possible. Therefore, in the case of complex-valued Kohn-Sham orbitals in the DFT calculations with plane-wave basis sets, we use an improved weighted function for the transposed Khatri-Rao product of two different sets of complex-valued Kohn-Sham orbitals defined as
\begin{equation}
w(\mathbf{r}) =  (\sum_{i=1}^{N_e}|\phi_i|^\alpha) + (\sum_{j=1}^{N}|\psi_j|^\alpha)
\label{eq:weight function 2}
\end{equation}
when the weight function $\alpha = $ 2.0 is the sum of the square modulus (abbreviated as SSM) of Kohn-Sham orbitals. It should be noticed that the SSM weight function forms the second term of wavefunction and the PSM weight function forms the fourth term of the wavefunction, because of the value of weight function is less than 1, such improved weight function is an alternative weight function that could show better results, superior to the PSM weight function defined in Eq.\ref{eq:weight function}. The $\alpha =$ 1.0, 3.0, and 4.0 represent that the weight function is the sum of the modulus (SM), cubic modulus (SCM) and quartic modulus (SQM) of Kohn-Sham orbitals, respectively. Eq.\ref{eq:weight function 2} is also valid for real-valued Kohn-Sham orbitals because real numbers are subsets of complex numbers. FIG.~\ref{fig:ponits} demonstrates the IPs selected by the K-means with SSM and PSM as well as QRCP procedures for BH$_3$NH$_3$ with different B-N distances. It is obvious that the K-means with SSM can yield IPs which are more dispersed and delocalized than the K-means with PSM and QRCP, which demonstrates that SSM is more suitable to simulate electron density as a weight function. In addition, we verify the feasibility of SSM as the weight function by demonstrating that the interpolation points using K-means with SSM approximately minimize the residual for the ISDF decomposition (See APPENDIX A).

\begin{algorithm}[H]
\caption{K-means Clustering Algorithm to Compute Interpolation Points in ISDF}

\leftline{\textbf{Input:} Grid points $\{\mathbf{r}_{i}\}_{i=1}^{N_{r}}$, Weight function $w(\mathbf{r})$.}

\leftline{\textbf{Output:} Interpolation points $\{\mathbf{r}_{\mu}\}_{\mu=1}^{N_{\mu}}$}

\begin{algorithmic}[1]

\STATE Initialize centroids $\{\mathbf{r}_{\mu}^{\{0\}}\}$, set $t \gets 0$.

%\STATE t \gets 0.

\WHILE {convergence not reached}

\STATE Classification step:  
           Assign $N_r$ points $\{\mathbf{r}_i\}_{i=1}^{N_{r}}$ to the cluster $C_{\mu}^{(t)}$

\STATE Compute new centroids: \\
$\mathbf{r}_{\mu}^{(t+1)} \gets {\sum_{\mathbf{r}_j \in C_{\mu}^{(t)}}\mathbf{r}_j w(\mathbf{r}_j)}/{\sum_{\mathbf{r}_j \in C_{\mu}^{(t)}}w(\mathbf{r}_j)}$

\STATE set $t \gets t + 1$

\ENDWHILE

\STATE Update $\{\mathbf{r}_{i}\}_{i=1}^{N_{\mu}} \gets \{\mathbf{r}_{\mu}^{\{t\}}\}$ .
\end{algorithmic} \label{alg:Kmeans}
\end{algorithm}

%\WH{Add the IPs for real and complex wavefunctions for QRCP and Kmeans for BH3-NH3, ask for Xinming Qin for details}

%\WH{Add the computational scaling and memory usage of real and complex wavefunctions for QRCP and Kmeans}

%\WH{Add the SVD single values of real and complex wavefunctions for QRCP and Kmeans}

\subsection{Low rank approximation of complex-valued hybrid DFT via ISDF}\label{sec:ISDFforHF}

The key spirit of DFT is to solve the Kohn-Sham equations expressed as
\begin{equation}
H \psi_j = (-\frac{1}{2}\Delta_{\mathbf{r}} + V_\text{ion} + V_\text{H}[\rho] + V_\text{XC}[\{\phi_i\}])\psi_j = \epsilon_j\psi_j  \label{eq:KS}
\end{equation}
where $H$ is the Kohn-Sham Hamiltonian, $\psi_j$ is the
$j$-th Kohn-Sham orbital, $\epsilon_j$ is the corresponding
orbital energy and $V_\text{ion}$ is the ionic potential. In real space, the Hartree
potential $V_\text{H}$ is defined as
\begin{equation}
V_\textrm{H}[\rho](\mathbf{r})=\int {\dfrac{\rho(\mathbf{r^{\prime}})}{|\mathbf{r}-\mathbf{r^{\prime}}|}}d\mathbf{r^{\prime}} \label{eq:hatp}
\end{equation}
the electron density is given as
\begin{equation}
\rho(\mathbf{r}) = \sum_{i=1}^{N_e}|\psi_i(\mathbf{r})|^2 
\label{eq:rho}
\end{equation}

It should be noticed that the accuracy of the Kohn-Sham DFT calculations strongly depends on the exchange-correlation potential $V_{XC}[\{\phi_i\}]$, which is defined as
\begin{equation}
V_{XC}[\{\phi_i\}] = V_{X}[\{\phi_i\}] + V_{C}[\rho]
\label{eq:vxc}
\end{equation}
where $\{\phi_i\}$ denote the occupied orbitals. $V_{X}[\{\phi_i\}]$ and $V_{C}[\rho]$ represent the exchange and correlation potentials, respectively. For complex-valued hybrid DFT, the Hartree-Fock exchange operator is
\begin{equation}
V_{X}[ \{ \phi_i \} ](\mathbf{r}, \mathbf{r^\prime}) = -\sum_{i=1}^{N_e} { \frac{\phi_i^\ast(\mathbf{r})\phi_i(\mathbf{r^\prime})} {|\mathbf{r}-\mathbf{r^\prime}|}}
\label{eq:vx}
\end{equation}

When the Hartree-Fock exchange operator is applied to the orbitals
\begin{equation}
(V_{X}[\{\phi_i\}]\psi_j)(\mathbf{r}) = -\sum_{i=1}^{N_e} \phi_i(\mathbf{r}) \int {\frac{\phi_i^\ast(\mathbf{r^\prime})\psi_j(\mathbf{r^\prime})}{|\mathbf{r}-\mathbf{r^\prime}|}{d}\mathbf{r^\prime}}
\label{eq:v}
\end{equation}

For large basis sets for discretizing the Kohn-Sham equations, such as the plane-wave basis sets, it is more efficient to use an iterative diagonalization procedure to compute the eq.~\ref{eq:KS}. In practice, several DFT packages, such as Quantum Espresso~\cite{giannozzi2009quantum} and PWDFT~\cite{hu2017adaptively}, separate the self-consistent field (SCF) iteration of all occupied orbitals into inner SCF iteration and outer SCF iteration, called two-level SCF produce. In the inner SCF iteration, the exchange operator $V_X$ defined by occupied orbitals $\{\phi_i\}$ in eq.~\ref{eq:vx} is fixed, so that the Hamiltonian operator only relies on the electron density $\rho$, which has to be updated constantly. In the outer SCF iteration, the output orbitals from the inner SCF iteration will be used for updating the exchange operator until it converges. In each inner SCF iteration, we must solve the two-center integrals of eq.~\ref{eq:v}. The practical numerical solution is to solve $O(N_e^2)$ Poisson-like equations, which is the most expensive part for hybrid DFT calculations within plane-wave basis sets. 

Under the ISDF decomposition of transposed Khatri-Rao product of complex-valued Kohn-Sham orbitals $Z =\{z_{ij}:= \phi_i^\ast(\mathbf{r})\psi_j(\mathbf{r})\} \in \mathbb{C}^{N_r
\times N_{e}^2}$, we can substitute Eq.(\ref{eq:decomposition}) into Eq.(\ref{eq:v})
\begin{small}
\begin{equation}
\begin{split}
(V_{X}[\{\phi_i\}]\psi_j)(\mathbf{r}) 
& \approx -\sum_{i=1}^{N_e} \phi_i(\mathbf{r}) \int {\frac{\sum_{\mu  =1}^{N_{\mu}}\zeta_{\mu}(\mathbf{r^\prime})\phi_i^\ast(\mathbf{r_{\mu}})\psi_j(\mathbf{r_{\mu}})}{|\mathbf{r}-\mathbf{r^\prime}|}{d}\mathbf{r^\prime}} \\
&= -\sum_{\mu =1}^{N_{\mu}} ({\int {\frac{\zeta_{\mu}(\mathbf{r^\prime})}{|\mathbf{r}-\mathbf{r^\prime}|}}{d}\mathbf{r^\prime} {{\sum_{i=1}^{N_e}\phi_i(\mathbf{r})\phi_i^\ast(\mathbf{r_{\mu}})\psi_j(\mathbf{r_{\mu}})}}})  \\
&= -\sum_{\mu =1}^{N_{\mu}} ({ V_{\mu}(\mathbf{r})
{{\sum_{i=1}^{N_e}\phi_i(\mathbf{r})\phi_i^\ast(\mathbf{r_{\mu}})\psi_j(\mathbf{r_{\mu}})}}})
\end{split}
\label{eq:v2}
\end{equation}
\end{small}
where the projected Hartree-Fock exchange integral under the ISDF decomposition is defined as
\begin{equation}
\begin{split}
V_{\mu}(\mathbf{r}) = \int {\frac{\zeta_{\mu}(\mathbf{r^\prime})}{|\mathbf{r}-\mathbf{r^\prime}|}}{d}\mathbf{r^\prime}
\end{split}
\label{eq:V_mu}
\end{equation}
As a consequence, the number of Poisson-like equations to be solved is reduced to $N_\mu \sim O(N_e)$ from $O(N_e^2)$.

\begin{figure*}[!htb]
\begin{center}
\includegraphics[width=1.00\textwidth]{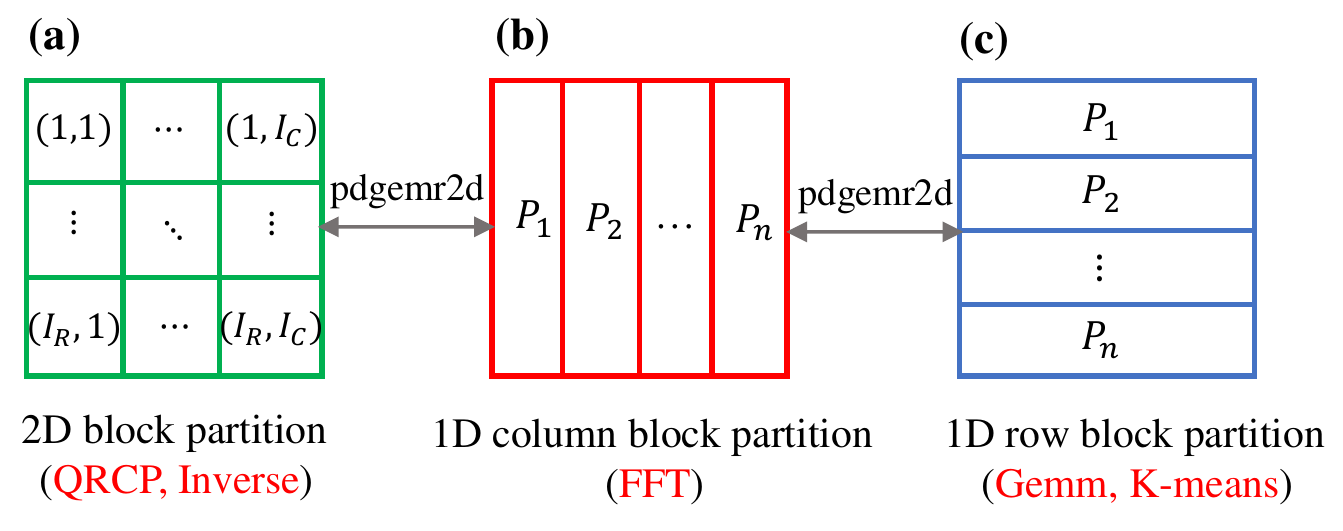}
\end{center}
\caption{Three different types of data partition for the matrix used in
the ISDF formulation for hybrid density functional calculations: (a) 2D
block cyclic partition ($I_R \times I_C$ MPI processor grid), (b) 1D column cyclic partition (1 × $P_n$ MPI processor grid), and (c) 1D row
cyclic partition ($P_n$ × 1 MPI processor grid). $P_n$ is total computational cores and $I_R \times I_C = P_n$.  } \label{fig:pdgemr2d}
\end{figure*}

\subsection{Parallel implementation}\label{sec:Parallel}

\begin{figure*}[!htb]
\begin{center}
\includegraphics[width=1.0\textwidth]{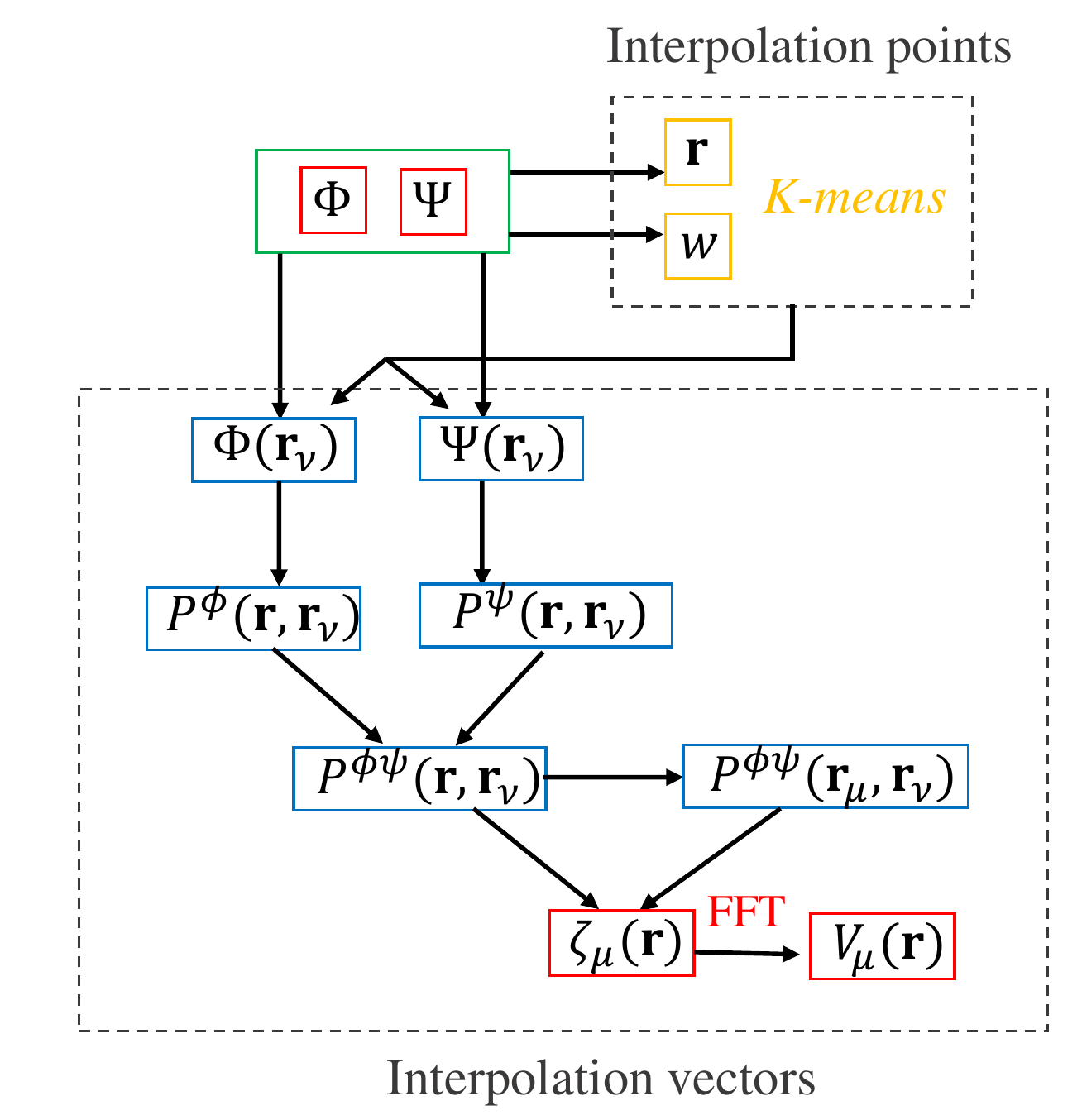}
\end{center}
\caption{Flowchart of the ISDF formulation in PWDFT. Red, blue and orange boxes denote 1D column block cyclic partition, 2D block cyclic partition and 1D row block cyclic partition, respectively.} \label{fig:parallel}
\end{figure*}

We implement this complex-valued K-means clustering algorithm for large-scale hybrid DFT calculations within plane-wave basis sets in PWDFT,~\cite{hu2017adaptively} which is an open source plane-wave based electronic structure calculations software. We also realize a parallel implementation of such low-scaling hybrid DFT calculations in PWDFT as shown in FIG.~\ref{fig:parallel}. 

The discretized Kohn-Sham orbitals can be denoted by $\Phi=[\phi_1, \phi_2,...,\phi_{N_e}]  \in \mathbb{C}^{N_r \times N_{e}}$ and $\Psi=[\psi_1, \psi_2,...,\psi_{N}]  \in \mathbb{C}^{N_r \times N}$. Thus the parallel is implemented easily with the aid of ScaLAPACK library when $P_n$ processors are used. There are three different data partition types including 1D column cyclic partition, 1D row cyclic partition and 2D block cyclic partition in our program, as shown in FIG.~\ref{fig:pdgemr2d}. It is easy for parallel implementation when we apply the Hamiltonian operator to the orbitals and use a sequential fast Fourier transformation (FFT) library by the 1D column block cyclic partition. The 2D block partition is suitable for QRCP method and matrix inversion, while the 1D row block partition should be adopted for matrix multiplication and K-means method. The transform among the different data partition types is achieved by the pdgemr2d subroutine in the ScaLAPACK library. FIG.~\ref{fig:parallel} demonstrates the flowchart of the ISDF formulation in PWDFT. Firstly, the orbitals $\Phi$ and $\Psi$ can be stored using the 1D column cyclic partition as the input. The interpolation points can be computed by the K-means clustering algorithm. The weight functions are computed from $\Phi$ and $\Psi$. The $N_r$ grid points are equally distributed to each core to compute the distances between grid points and centroids for parallel implementation of the K-means clustering part. Then the quasi density matrices $P^{\phi}(\mathbf{r}_{\mu},\mathbf{r}_{\nu}) \in \mathbb{C}^{N_{\mu} \times N_{\mu}}$ and $P^{\psi}(\mathbf{r}_{\mu},\mathbf{r}_{\nu})  \in \mathbb{C}^{N_{\mu} \times N_{\mu}} $ in Eq.~\ref{eq:muRownuColden} are transformed into the 2D block cyclic partition to construct the IVs. The matrix $ZC^T$ in Eq.~\ref{eq:theta} can be calculated in parallel fashion and the matrix $CC^T$ is exactly subsampling rows of $ZC^T$. After we obtain the IVs $\Theta = [\zeta_1, \zeta_2,...,\zeta_{N_\mu}] \in \mathbb{C}^{N_r \times N_{\mu}} $ by linear equation solver in ScaLAPACK, the data partition type should be converted from 2D block cyclic partition to 1D column cyclic partition for computing the $V = [V_1, V_2,...,V_{N_\mu}] \in \mathbb{C}^{N_r \times N_{\mu}}$.

\begin{table*}[!htb] \footnotesize
\begin{center}
\setlength{\tabcolsep}{5mm}
\caption{Accuracy of complex-valued ISDF (SSM, $\alpha = 2$) based HSE06 calculations by using K-means clustering algorithm to compute the IPs with respect to the rank parameter $t$ for liquid water molecules (H$_2$O)$_{64}$, semiconducting solid Si$_{216}$ and metallic aluminium-silicon alloy Al$_{176}$Si$_{24}$, including the VBM $E_\textrm{VBM}$ (eV), CBM $E_\textrm{CBM}$ (eV), the energy gap $E_\textrm{g}$ (eV), the absolute errors of Hartree-Fock exchange energy $\Delta E_\textrm{HF}$ (Ha/atom), total energy $\Delta E$ (Ha/atom) and atomic forces $\Delta F$ (Ha/Bohr). The ACE-enabled HSE06 calculations are used for the reference.}
\label{tab:kmeans_accuracy}
\begin{threeparttable}
\begin{tabular}{ccccccc}
\hline \hline
$t$  &$E_\textrm{VBM}$ &$E_\textrm{CBM}$ &$E_\textrm{g}$ &$\Delta E_\textrm{HF}$ &$\Delta E$  &$\Delta F$  \\
\hline
 \multicolumn{7}{c}{Liquid water molecules (H$_2$O)$_{64}$ ($N_\textrm{band} = 255$)}\\
  \hline 
4.0  &-3.8136  &2.4196  &6.2333  &$2.22\times 10^{-04}$ &$3.12\times 10^{-04}$ &$1.30\times 10^{-02}$\\ 
6.0  &-3.8261  &2.2910  &6.1170  &$2.63\times 10^{-05}$ &$4.39\times 10^{-05}$ &$5.59\times 10^{-04}$\\ 
8.0  &-3.8291  &2.2740  &6.1031  &$3.46\times 10^{-06}$ &$1.49\times 10^{-05}$ &$3.22\times 10^{-04}$\\ 
10.0 &-3.8298  &2.2713  &6.1011  &$1.26\times 10^{-07}$ &$1.86\times 10^{-05}$ &$6.54\times 10^{-05}$\\ 
12.0 &-3.8301  &2.2708  &6.1010  &$5.68\times 10^{-07}$ &$2.81\times 10^{-05}$ &$5.01\times 10^{-05}$\\
16.0 &-3.8302  &2.2706  &6.1008  &$6.65\times 10^{-07}$ &$2.87\times 10^{-05}$ &$5.68\times 10^{-05}$\\ 
20.0 &-3.8300  &2.2706  &6.1006  &$3.65\times 10^{-09}$ &$4.17\times 10^{-07}$ &$2.11\times 10^{-05}$\\ 
Ref  &-3.8299  &2.2705  &6.1004  &$0.00\times 10^{00}$  &$0.00\times 10^{00}$  &$0.00\times 10^{00}$\\
 \hline 
  \multicolumn{7}{c}{Semiconducting bulk silicon solid Si$_{216}$ ($N_\textrm{band} = 432$)}\\
   \hline 
4.0  &6.7419  &8.3454  &1.6035  &$2.53\times 10^{-03}$ &$2.96\times 10^{-03}$ &$2.73\times 10^{-03}$\\ 
6.0  &6.6641  &8.1544  &1.4903  &$3.66\times 10^{-04}$ &$4.64\times 10^{-04}$ &$6.47\times 10^{-04}$\\ 
8.0  &6.6513  &8.1040  &1.4527  &$7.45\times 10^{-05}$ &$9.97\times 10^{-05}$ &$1.82\times 10^{-04}$\\ 
10.0 &6.6480  &8.0959  &1.4479  &$1.87\times 10^{-05}$ &$2.83\times 10^{-05}$ &$1.99\times 10^{-04}$\\ 
12.0 &6.6472  &8.0942  &1.4470  &$4.73\times 10^{-06}$ &$9.66\times 10^{-06}$ &$4.56\times 10^{-05}$\\
16.0 &6.6469  &8.0932  &1.4463  &$1.77\times 10^{-07}$ &$1.51\times 10^{-06}$ &$1.60\times 10^{-05}$\\ 
20.0 &6.6468  &8.0930  &1.4462  &$4.27\times 10^{-07}$ &$3.66\times 10^{-07}$ &$5.35\times 10^{-06}$\\ 
Ref  &6.6468  &8.0930  &1.4462  &$0.00\times 10^{00}$  &$0.00\times 10^{00}$  &$0.00\times 10^{00}$\\
\hline
  \multicolumn{7}{c}{Metallic aluminium-silicon alloy Al$_{176}$Si$_{24}$ ($N_\textrm{band} = 312$)}  \\
   \hline 
4.0  &7.9370  &8.0369  &0.0999  &$3.93\times 10^{-03}$ &$4.20\times 10^{-03}$ &$4.09\times 10^{-03}$\\ 
6.0  &7.8118  &7.9168  &0.1050  &$6.91\times 10^{-04}$ &$7.29\times 10^{-04}$ &$1.22\times 10^{-03}$\\ 
8.0  &7.7760  &7.8759  &0.0999  &$8.50\times 10^{-05}$ &$8.82\times 10^{-05}$ &$3.20\times 10^{-04}$\\ 
10.0 &7.7714  &7.8703  &0.0989  &$1.76\times 10^{-05}$ &$1.88\times 10^{-05}$ &$1.02\times 10^{-04}$\\ 
12.0 &7.7708  &7.8695  &0.0987  &$6.13\times 10^{-06}$ &$7.04\times 10^{-06}$ &$4.82\times 10^{-05}$\\
16.0 &7.7706  &7.8693  &0.0986  &$1.09\times 10^{-06}$ &$1.77\times 10^{-06}$ &$2.96\times 10^{-05}$\\ 
20.0 &7.7705  &7.8691  &0.0986  &$1.73\times 10^{-07}$ &$6.25\times 10^{-07}$ &$1.18\times 10^{-05}$\\ 
Ref  &7.7705  &7.8691  &0.0986  &$0.00\times 10^{00}$  &$0.00\times 10^{00}$  &$0.00\times 10^{00}$\\
%\multirow{2}{*}{C6H6} &6.0          &0.056     &$1.65\times 10^{-2}$      &$2.12\times 10^{-2}$      
\hline \hline
\end{tabular}
%\begin{tablenotes}
%\footnotesize
%\end{tablenotes}
\end{threeparttable}
\end{center}
\end{table*}

\section{Results and discussion} \label{sec:Results}

In this section, we demonstrate the numerical accuracy and computational efficiency of the K-means clustering algorithm of complex-valued Kohn-Sham orbitals for ISDF to accelerate hybrid density functional calculations. All calculations are implemented in the PWDFT~\cite{hu2017adaptively} software package and Message Passing Interface (MPI) is used for handling data communication. We use the Hartwigsen-Goedecker-Hutter (HGH) norm-conserving pseudopotentials\cite{1998Relativistic} and the HSE06~\cite{2006Erratum}  functional to describe the electronic structures of molecules and solids.
\begin{figure*}[!htb]
\begin{center}
\includegraphics[width=1.00\textwidth]{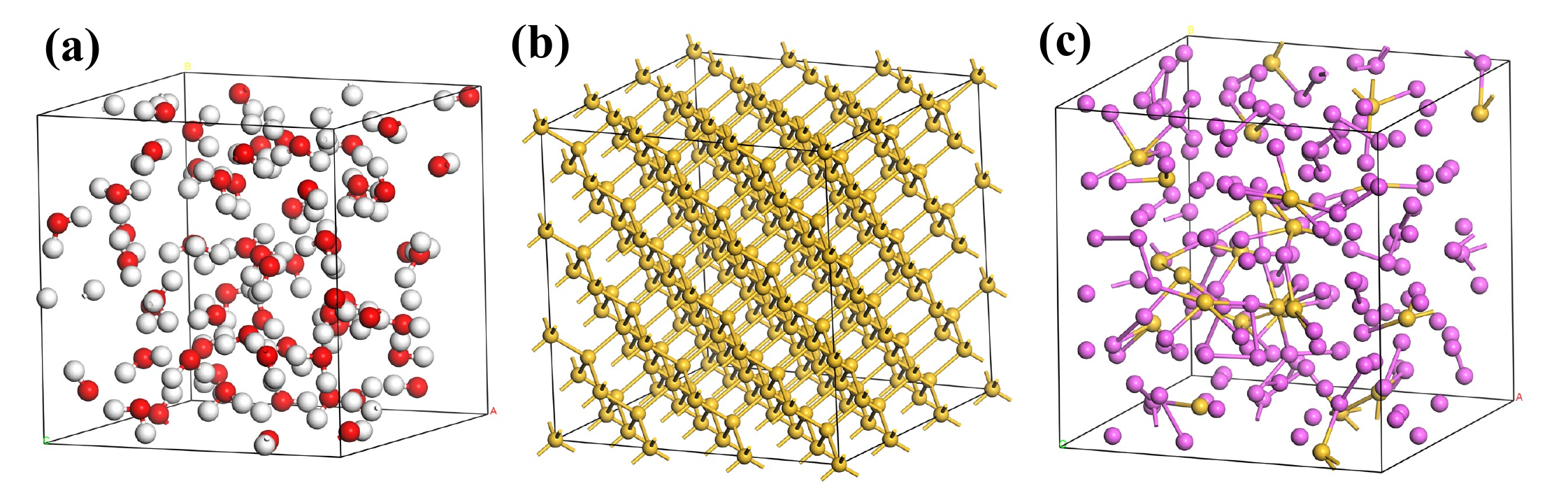}
\end{center}
\caption{Atomic structures of (a) insulator liquid water molecules (H$_2$O)$_{64}$, (b) semiconducting bulk silicon solid Si$_{216}$ and (c) metallic aluminium-silicon alloy Al$_{176}$Si$_{24}$. The red, white, yellow and purple circles denote O, H, Si and Al atoms, respectively.} \label{fig:structure}
\end{figure*}

We first benchmark the numerical accuracy of ISDF-K-means with weight function SSM by comparing standard HSE06 results. Then we compare different numerical accuracy of ISDF-K-means with SSM, ISDF-K-means with PSM and ISDF-QRCP. Furthermore, we compare the AIMD results from K-means with different weight functions. Finally, we demonstrate the computational scaling of ISDF-K-means as well as ISDF-QRCP and the parallel scalability of ISDF-enabled hybrid density functional calculations on modern heterogeneous supercomputers.

\subsection{Numerical accuracy}\label{sec:Accuracy} 

\subsubsection{Total energy and atomic forces}\label{sec:Energy} \ \\

Firstly, we test the accuracy of the ISDF method for complex-valued Kohn-Sham orbitals taking liquid water molecules (H$_2$O)$_{64}$, semiconducting bulk silicon Si$_{216}$ ($E_{gap}$ = 1.45 eV) and metallic disordered aluminium-silicon alloy Al$_{176}$Si$_{24}$ ($E_{gap}$ $\textless$ 0.1 eV) as examples, whose crystal structures are demonstrated in FIG.~\ref{fig:structure}. The cutoff energies for (H$_2$O)$_{64}$, Si$_{216}$ and Al$_{176}$Si$_{24}$ are 60.0, 20.0 and 20.0 Ha, respectively. For (H$_2$O)$_{64}$ system, the DFT-D2 method is used to account for the van der Waals (VdW) interaction.~\cite{grimme2006semiempirical} We obtain all reference results utilizing adaptively compressed exchange (ACE) algorithm,~\cite{lin2016adaptively,hu2017adaptively} which reduces the cost of applying Hartree-Fock exchange operator into Kohn-Sham orbitals without loss of accuracy.
\begin{figure*}[!htb]
\begin{center}
\includegraphics[width=0.82\textwidth]{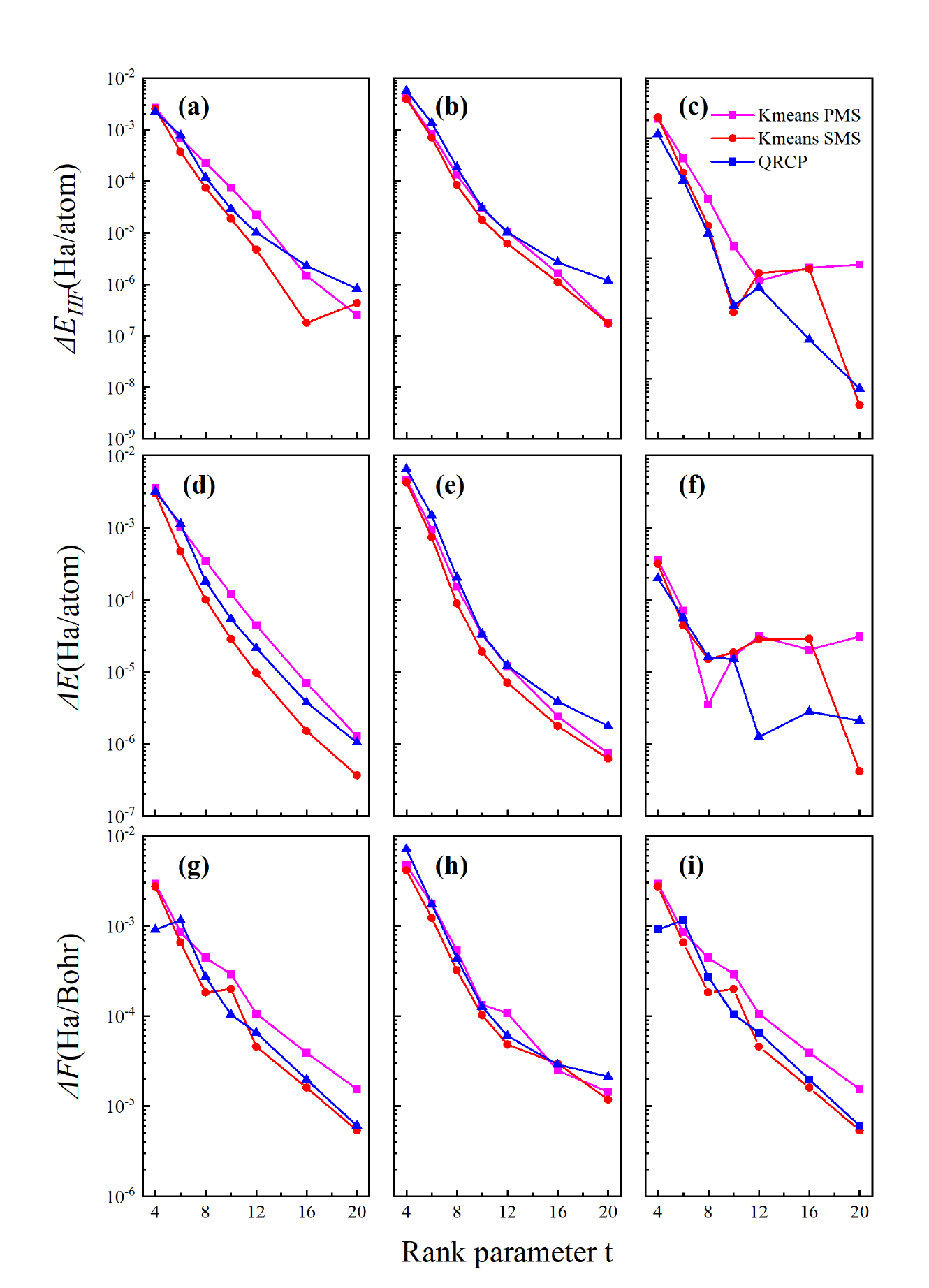}
\end{center}
\caption{Accuracy of complex-valued ISDF based hybrid functional calculations
(HSE06) obtained by using the K-means (SMS and PMS) and QRCP procedures to select
the interpolation points, with varying rank parameter t from 4 to 20
for Si$_{216}$, Al$_{176}$Si$_{24}$ and (H$_2$O)$_{64}$, including the error of ((a), (b), (c)) Hartree-Fock exchange energy $\Delta E_\textrm{HF}$ (Ha/atom), ((d), (e), (f)) total energy $\Delta E$ (Ha/atom) and ((g), (h), (i)) atomic forces $\Delta F$ (Ha/Bohr).} \label{fig:EtotEhfF}
\end{figure*}

Table~\ref{tab:kmeans_accuracy} demonstrates the valence band maximum (VBM) energy level, the conduction band minimum (CBM) energy level, the energy gap $E_\textrm{g}$, the absolute errors of Hartree-Fock exchange energy $\Delta E_\textrm{HF}$, total energy $\Delta E$ (Ha/atom) as well as atomic forces $\Delta F$ (Ha/Bohr) by ISDF-based HSE06 calculations using K-means with weight function SMS with respect to the rank parameter $t$. Here we define 
\begin{equation}
\begin{aligned}
&\Delta E_\textrm{HF} = (E_\textrm{HF}^{ISDF}-E_\textrm{HF}^{Ref})/N_A \\
&\Delta E = (E^{ISDF}-E^{Ref})/N_A \\
&\Delta F = \max \limits_{I}|(F_I^{ISDF}-F_I^{Ref})|
\end{aligned}
\label{eq:delaE}
\end{equation}
where $N_A$ is the total number of atoms, and $I$ is the atom index. 
\begin{figure*}[!htb]
\begin{center}
\includegraphics[width=0.83\textwidth]{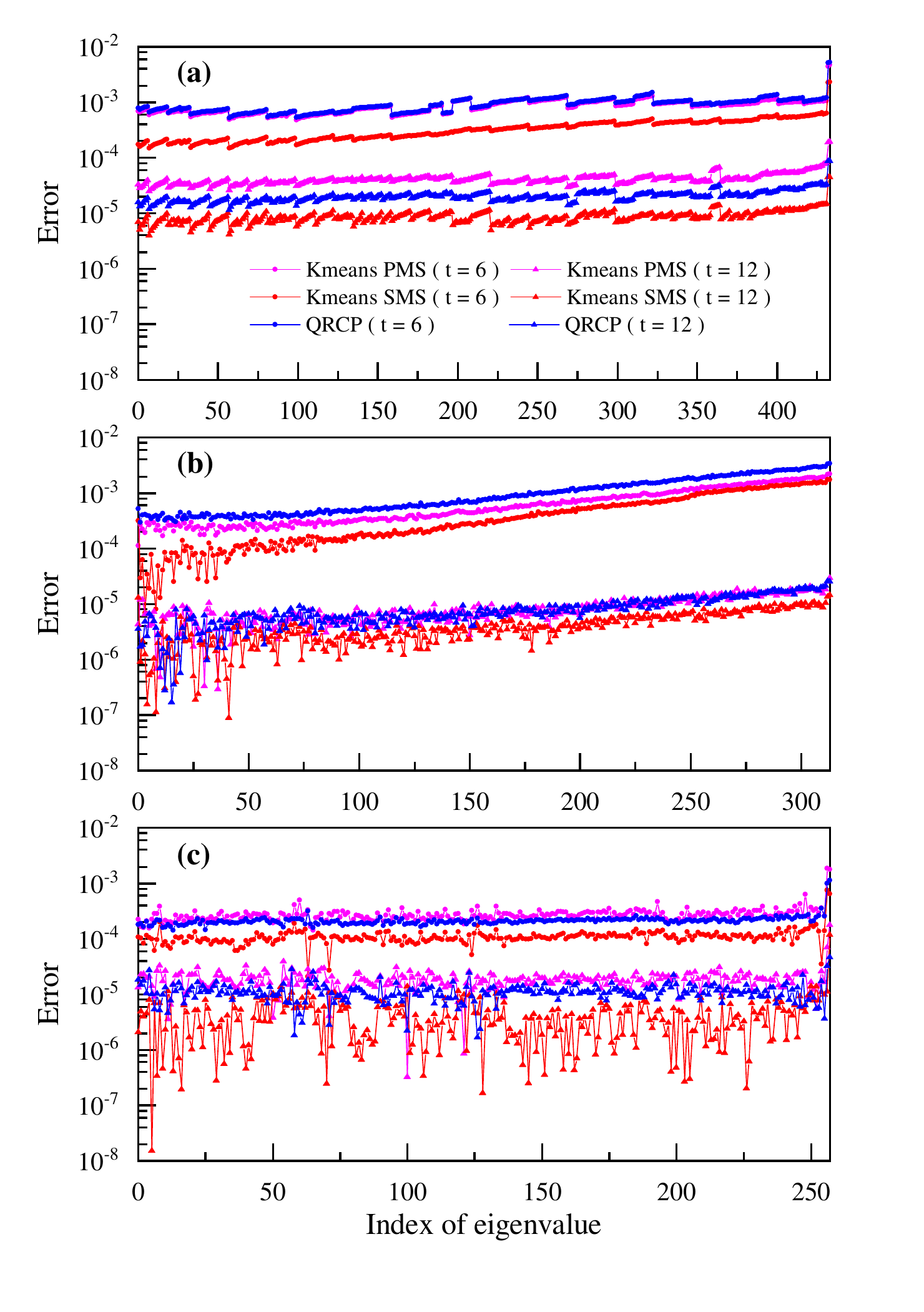}
\end{center}
\caption{Variation of eigenvalue error in hybrid HSE06 DFT calculations using K-means with SSM and PSM as well as QRCP method with respect to the rank parameter $t$ ($t$ = 6 and $t$ = 12) for (a) Si$_{216}$, (b) Al$_{176}$Si$_{24}$ and (c) (H$_2$O)$_{64}$.} \label{fig:eigenError}
\end{figure*}

We remark that above physical quantities are approaching the reference value gradually as the rank parameter $t$ is increasing for three studied systems. Besides, for all tested systems, the error of energy gap is less than 0.01 eV and the error of total energy per atom is less than $10^{-4}$ Ha/atom when $t$ is set to 8.0, which suggests that a small rank parameter can yield enough accurate results. The accuracy of standard method QRCP as well as K-means method with SSM and PSM are compared in FIG.~\ref{fig:EtotEhfF}. When the same rank parameter $t$ is set, the K-means method with SSM yields higher accuracy than K-means with PSM for Hartree-Fock exchange energy, total energy and atomic forces in most cases. In order to further compare the accuracy of K-means with different weight functions and QRCP, we demonstrate variation of eigenvalue error with different rank parameters $t$ = 6.0 and 12.0 for Si$_{216}$, Al$_{176}$Si$_{24}$ and (H$_2$O)$_{64}$, as shown in FIG.~\ref{fig:eigenError}. The error of the $i$-th eigenvalue is calculated by $\Delta \epsilon_i  = \epsilon_i^{ISDF} - \epsilon_i^{Ref}$. Similarly, it is obvious that the K-means with SSM shows a lower error of eigenvalues for all tested systems.

\subsubsection{AIMD}\label{sec:AIMD} \ \\

In order to further verify the accuracy of the ISDF method with the improved K-means algorithm, we perform AIMD optimized by hybrid DFT calculations for bulk silicon system Si$_{64}$, aluminium-silicon alloy Al$_{176}$Si$_{24}$ and silicon dioxide SiO$_2$ under the NVE ensemble and liquid water molecules (H$_2$O)$_{64}$ under the NVT ensemble. For NVT ensemble, a single level Nose-Hoover thermostat~\cite{nose1984unified,hoover1985canonical} is used at 295 K of temperature and the mass of the Nose-Hoover thermostat is 85,000 a.u.
\begin{figure*}[!htb]
\begin{center}
\includegraphics[width=1.00\textwidth]{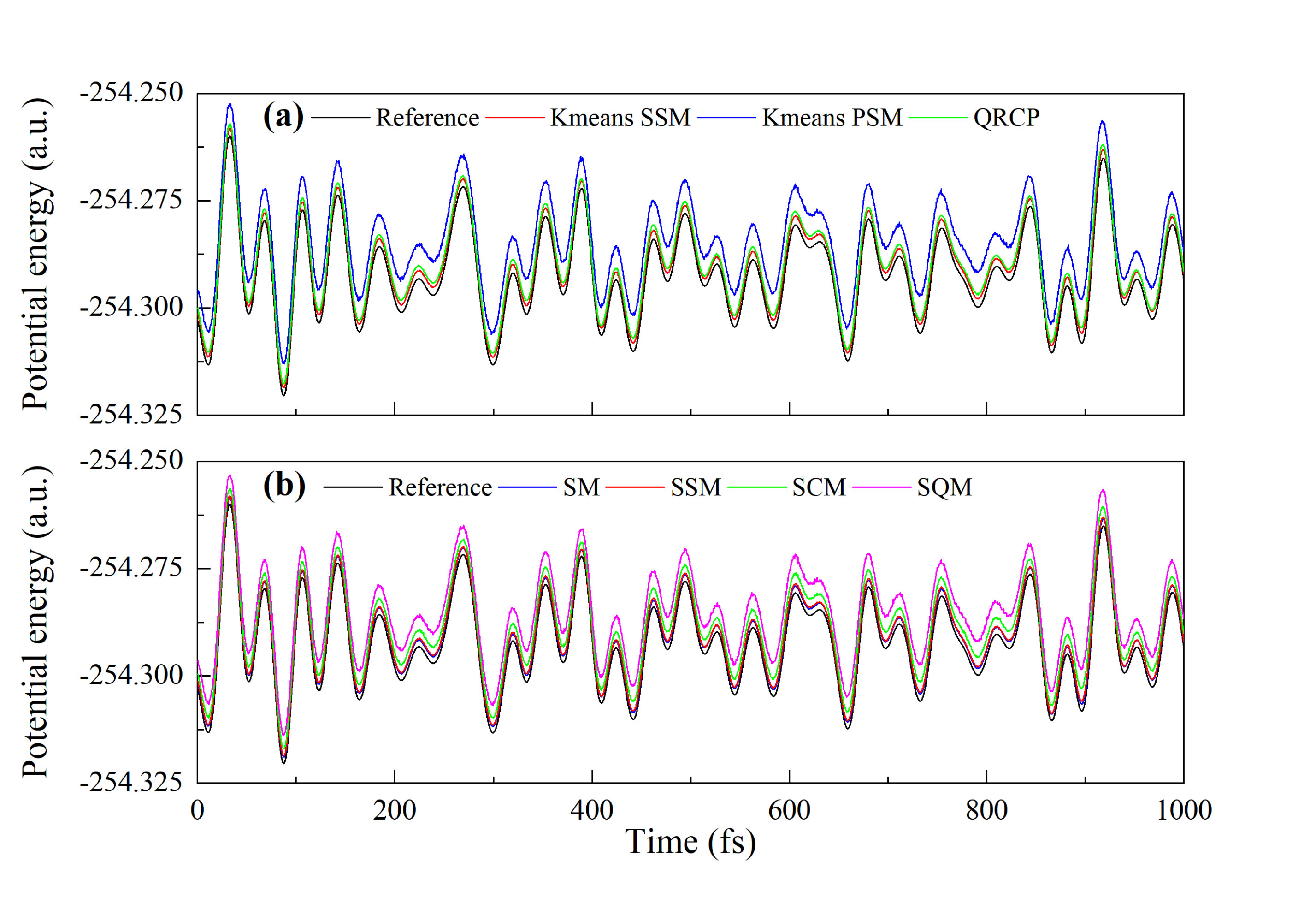}
\end{center}
\caption{Energy potential of hybrid HSE06 DFT AIMD simulations by using the ISDF-K-means with different weight functions and ISDF-QRCP methods as well as ACE-enabled procedure as the reference on the bulk silicon Si$_{64}$.} \label{fig:Si64_md_Epot}
\end{figure*}

FIG.~\ref{fig:Si64_md_Epot} demonstrates that ISDF-K-means with different weight functions and ISDF-QRCP method results in the total potential energy along the MD trajectory with a time step of 1.0 fs on bulk silicon system Si$_{64}$. We remark that the absolute error from K-means method with SSM is smaller than that from K-means method with PSM. Moreover, the K-means method with SSM yields a much smoother potential energy change compared with that yielded by the K-means method with PSM. The possible reason is that the interpolation points selected by weight function SSM represent the distribution of electron density better than that by weight function PSM. Therefore, weight function SSM will perform better in some cases where a smooth potential energy surface is required, such as geometry optimization. In addition, we also compare the potential energy change along the MD trajectory of K-means method with SM ($\alpha=1$), SSM ($\alpha=2$), SCM ($\alpha=3$) as well as SQM ($\alpha=4$). We remark that the weight function SM exhibits a similar performance to the weight function SSM. Note that the K-means method with SQM yields almost the same potential energy change as that from the K-means method with PSM. It is because both weight function SQM and PSM are essentially quartic modulus of wavefunctions in hybrid functional calculations where $\Psi$ is the same as $\Phi$ when the unoccupied orbitals are not computed. 
\begin{figure*}[!htb]
\begin{center}
\includegraphics[width=1.00\textwidth]{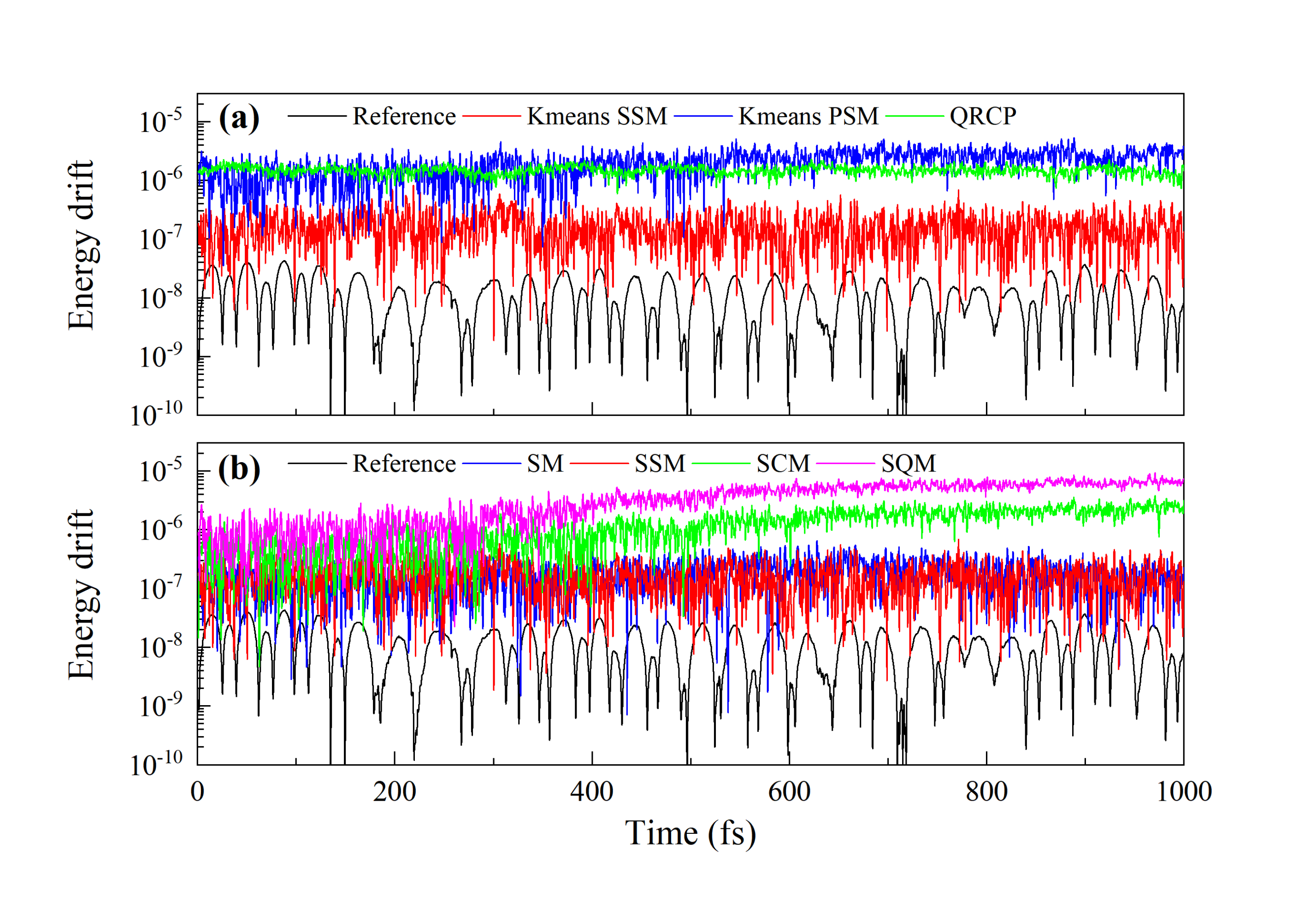}
\end{center}
\caption{Relatively energy drift of hybrid HSE06 AIMD simulations by using the ISDF-K-means with different weight functions and ISDF-QRCP methods as well as ACE-enabled procedure as the reference on the bulk silicon Si$_{64}$.} \label{fig:Si64_md_Edrift}
\end{figure*}

\begin{figure*}[!htb]
\begin{center}
\includegraphics[width=1.00\textwidth]{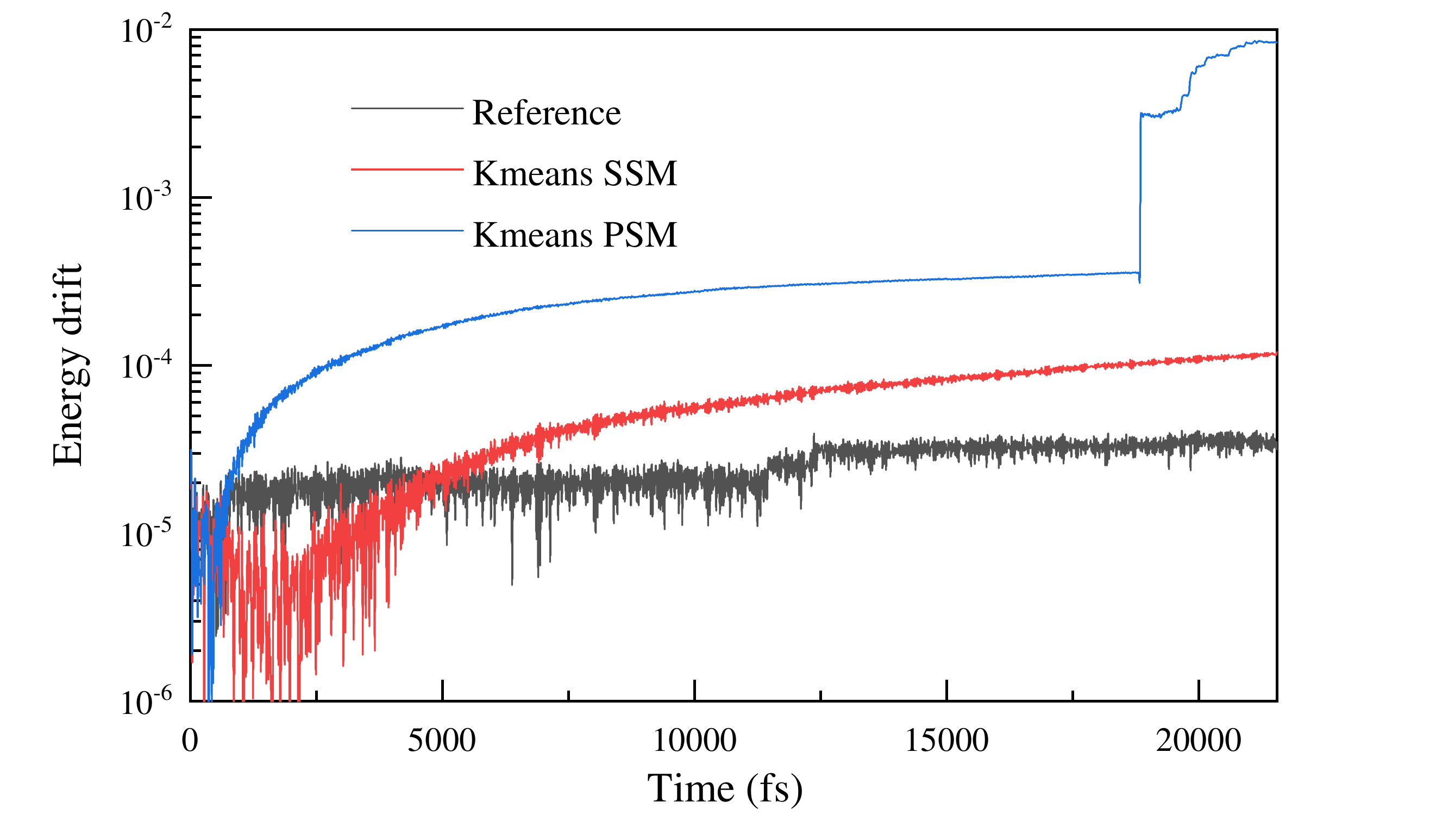}
\end{center}
\caption{Relatively energy drift of hybrid HSE06 AIMD simulations with a time step of 12.0 fs for Si$_{64}$.} \label{fig:longDT}
\end{figure*}

Energy drift represents the change of total energy compared to the initial total energy along the MD trajectory, which can be defined by $E_{drift}(t) = (E_{tot}(t)-E_{tot}(0))/E_{tot}(0)$. FIG.~\ref{fig:Si64_md_Edrift} shows the controlled energy drift by the ISDF-K-means with different weight functions and ISDF-QRCP on bulk silicon system Si$_{64}$. We remark that the K-means with SSM exhibits less loss of accuracy than the K-means with PSM and QRCP methods. Besides, divergence occurs for the energy drift of K-means with PSM while that of K-means with SSM maintains stable accuracy. On the other hand, the performance of the K-means method with SM is similar to that of the K-means method with SSM, while the performance of the K-means method with SQM is similar to that of the K-means method with PSM, which is consistent with the results from the energy potential change. In addition, for AIMD with a time step of 1.0 fs on aluminium-silicon alloy Al$_{176}$Si$_{24}$, as FIG.S1 and FIG.S2 show, the K-means method with SSM also yields slightly smoother potential energy and less loss of accuracy of energy drift than the K-means method with PSM. Similarly, for liquid water molecules, the K-means with SSM exhibits higher accuracy than the K-means with PSM on the whole, as shown in FIG.S5. From FIG.S4, the potential energy yielded by the K-means with SSM is more consistent with the reference results compared with that by the K-means with PSM for water molecules, which suggests a stronger stability of SSM. In order to further check the stability of the K-means with SSM method, we perform a long MD simulation for 21.5 ps with a time step of 12.0 fs on Si$_{64}$, as FIG.~\ref{fig:longDT} shows. K-means with SSM exhibits less loss of accuracy and better stability than K-means with PSM.

\begin{figure*}[!htb]
\begin{center}
\includegraphics[width=1.0\textwidth]{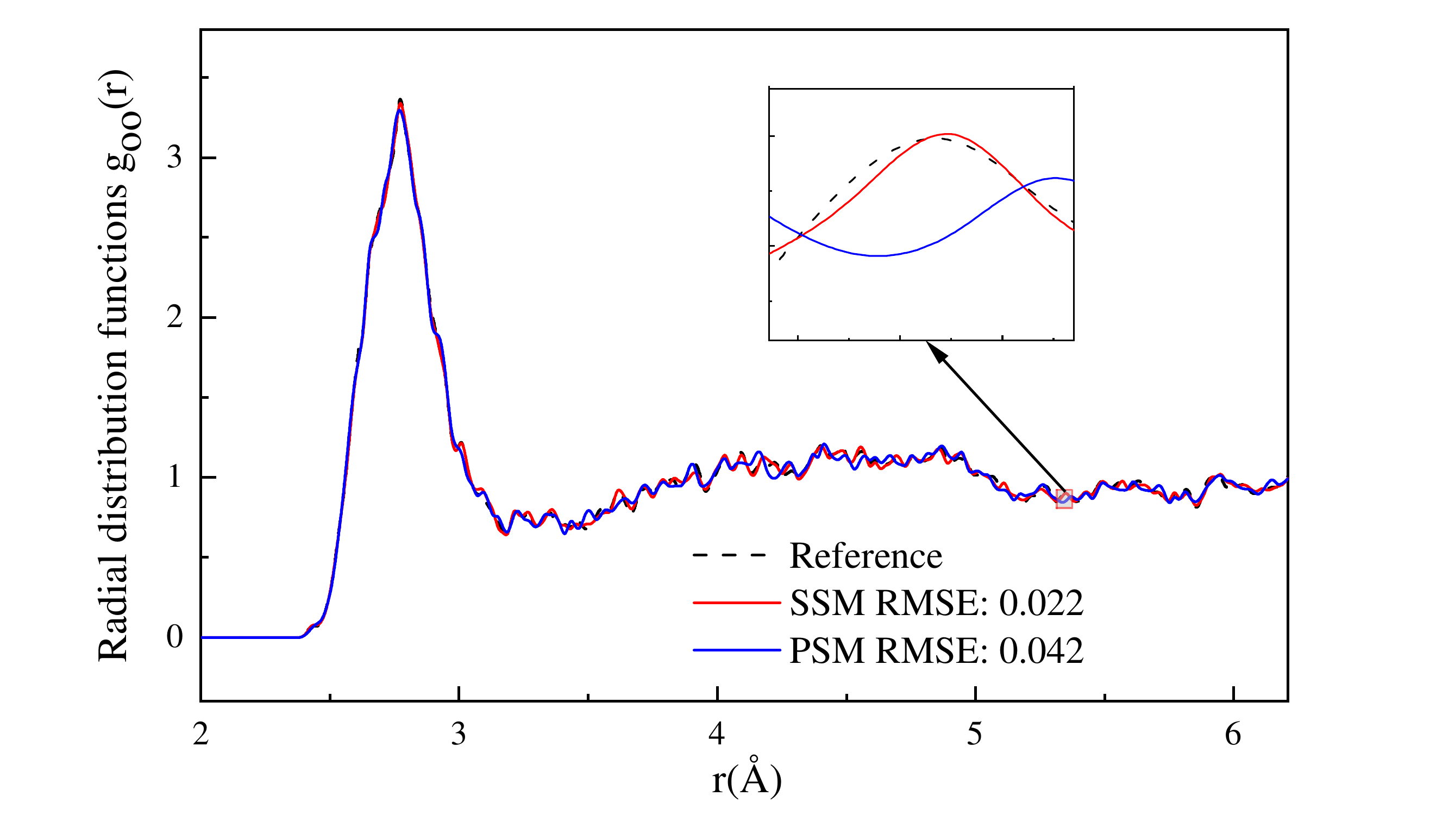}
\end{center}
\caption{Oxygen-oxygen radial distribution functions g$_{OO}$(r) of the liquid water system (H$_2$O)$_{64}$ at 295 K obtained from hybrid HSE06 + DFT-D2 AIMD simulations for 0.37 ps with a time step 1.0 fs using the ISDF-K-means with SSM (red solid line) and ISDF-K-means with PSM (blue solid line) methods and the ACE algorithm (black dashed line as the reference). The root mean square error (RMSE) of SSM and PSM with respect to the reference is 0.022 and 0.042, respectively.} \label{fig:rdf}
\end{figure*}

We also sample the oxygen-oxygen radial distribution functions (RDF) by hybrid HSE06 AIMD simulations for 0.37 ps with a time step of 1.0 fs on liquid water system (H$_2$O)$_{64}$ under NVT ensemble at 295 K using K-means with SSM and K-means with PSM, as FIG.~\ref{fig:rdf} shows. We remark that the root mean square error (RMSE) of K-means with SSM (0.022) is much smaller than that of K-means with PSM (0.042), which demonstrates better accuracy for SSM compared with PSM. Therefore, this improved K-means clustering algorithm in the ISDF decomposition can accurately and efficiently realize low-scaling, large-scale and long-time AIMD with complex-valued hybrid DFT calculations.

\begin{figure*}[!htb]
\begin{center}
\includegraphics[width=1.0\textwidth]{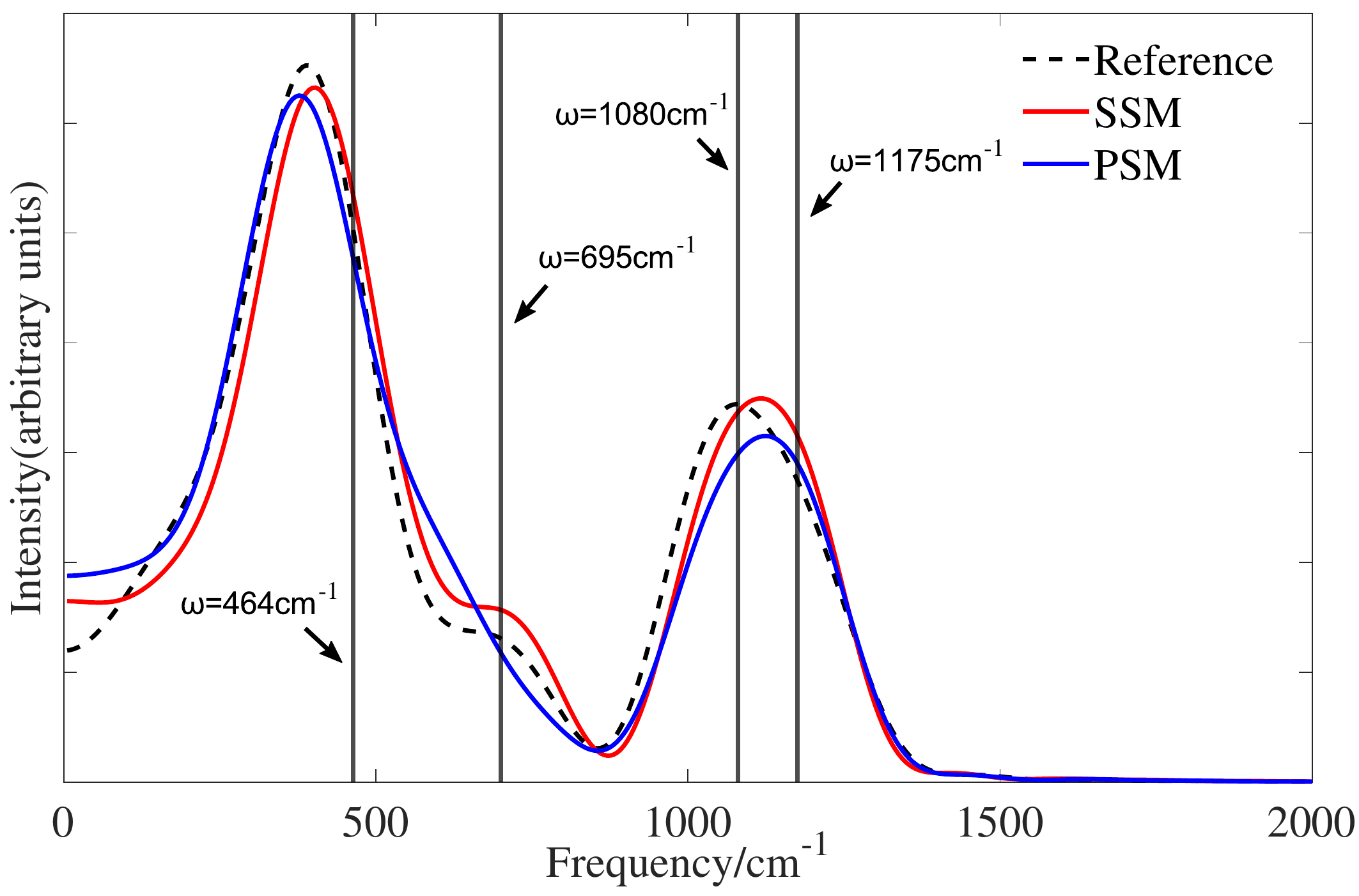}
\end{center}
\caption{Power spectrum of the crystal silicon dioxide system SiO$_2$ at 300K obtained from hybrid HSE06 AIMD simulations for 2ps with a time step 1.0 fs using the ISDF-K-means with SSM (red solid line) and ISDF-K-means with PSM (blue solid line) methods and the ACE algorithm (black dashed line as the reference). The experimental results are shown as four black vertical lines.} \label{fig:power}
\end{figure*}

In addition to studying the oxygen-oxygen RDF of liquid water system (H$_2$O)$_{64}$, we also calculate the power spectrum~\cite{spectral} of the crystal silicon dioxide SiO$_2$ by using HSE06 AIMD simulations under NVE ensemble at 300 K using K-means with SSM and K-means with PSM. As shown in FIG.~\ref{fig:power}, the black vertical line shows three different vibration frequencies in experiment~\cite{spectrumexp}, which is Si-O asymmetrical bending vibration with 464 cm $^{-1}$,  Si-O symmetrical bending vibration with 695 cm$^{-1}$ and Si-O asymmetrical stretching vibration with 1080-1175 cm$^{-1}$, respectively, we could see it is closer to the asymmetrical bending vibration frequency and asymmetrical bending vibration frequency when using K-means with SSM. Furthermore, we could find a small peak with 695 cm$^{-1}$ using K-means with SSM, however, we couldn't see this in K-means with PSM. Therefore, the improved K-means clustering with SSM in ISDF algorithm could simulate more accurate power spectral vibrational frequencies in hybrid AIMD simulations.

\subsection{Computational efficiency}\label{sec:Efficiency}  

\subsubsection{Computational scaling}\label{sec:Scaling}  \ \\

To compare the computational efficiency of K-means and QRCP method, we demonstrate the computational time of IPs selected by QRCP and K-means method and IVs for complex-valued wavefunctions in FIG.S5. The tested systems are bulk silicon systems including from 8 to 512 atoms. The cutoff energy is set to 10 Ha. From the fitting curves, we remark that selecting IPs by K-means is much faster than that by QRCP. For periodic systems where $N_r$ is proportional to the number of atoms, the K-means algorithm scales as $O(N^{2.0})$ while the QRCP method scales as $O(N^{3.0})$, which is consistent with the conclusion in numerical atomic orbitals.~\cite{qin2020machine} The IVs procedure by the least-square method scales as $O(N^{2.4})$. For ISDF-QRCP method, the IPs procedure is the most time-consuming part due to the expensive QRCP. However, the IVs procedure becomes the dominant part when the K-means method replaces the QRCP.

\subsubsection{Parallel scalability}\label{sec:Scalability} \ \\
\begin{figure*}[!htb]
\begin{center}
\includegraphics[width=1.0\textwidth]{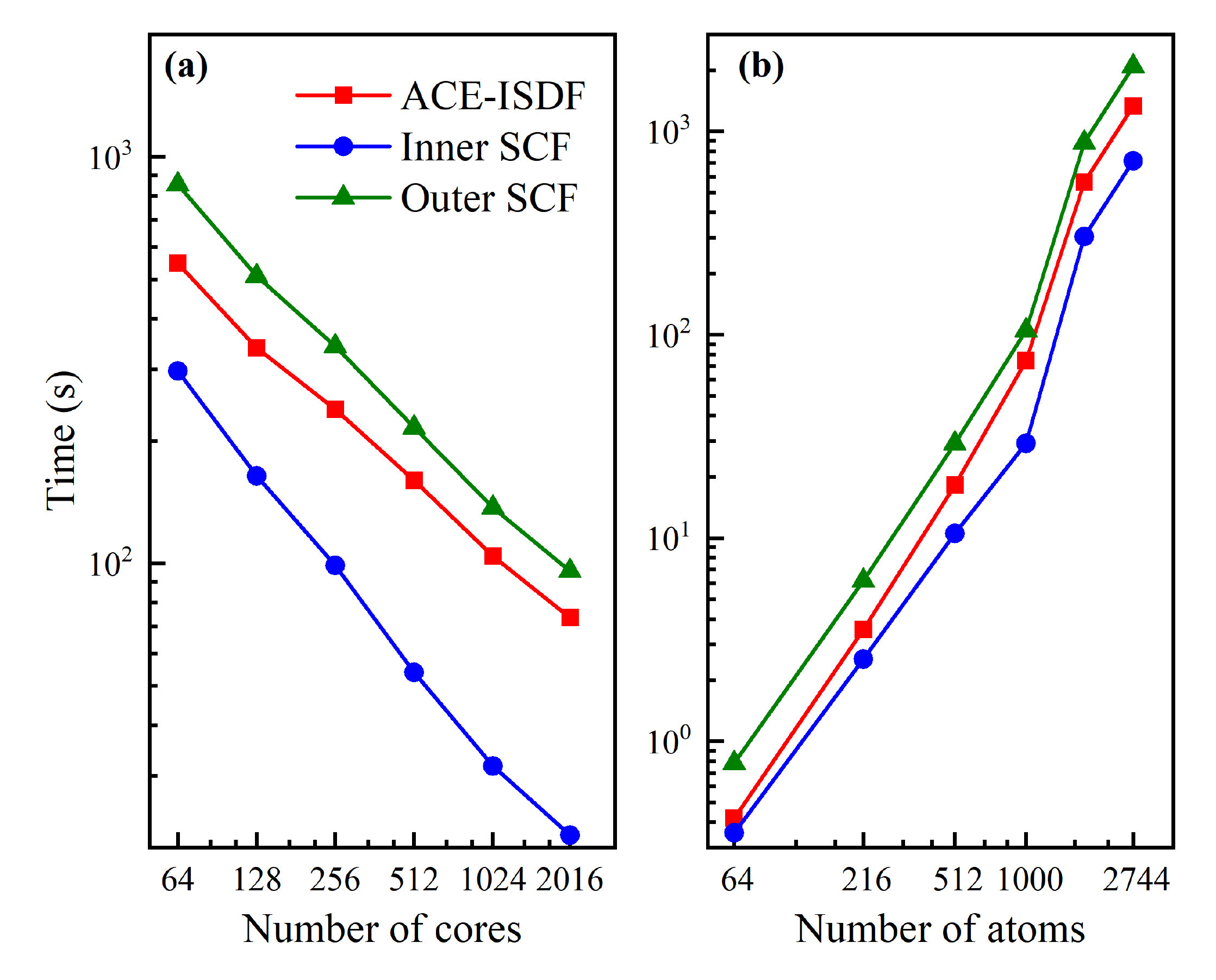}
\end{center}
\caption{(a) The change of wallclock time in ISDF, inner SCF and outer SCF iteration
with respect to the number of cores for the Si$_{1000}$ system. (b) The change of wallclock time in ISDF, inner SCF and outer SCF iteration with respect to system size.} \label{fig:pall}
\end{figure*}

We test the parallel scalability of the complex-valued ISDF method with K-means clustering algorithm for large-scale hybrid density functional (HSE06) calculations, as shown in FIG.~\ref{fig:pall}. FIG.~\ref{fig:pall}(a) shows the change of the wallclock time of ISDF part, inner SCF and outer SCF in one outer SCF iteration with respect to the numbers of cores for the bulk silicon system containing 1,000 atoms, which illustrates strong parallel scalability of our algorithm. We remark that the wallclock time of inner SCF, ISDF and outer SCF exhibits excellent scalability up to 2,000 CPU cores. As for the weak parallel scalability, FIG.~\ref{fig:pall}(b) demonstrates the change of wallclock time with respect to the numbers of atoms for the bulk silicon system including from 64 to 2,744 atoms. The ISDF method scales well with respect to the system size (up to 2,744 atoms) on 5,504 CPU cores. The hybrid density functional calculations for complex-valued Kohn-Sham orbitals require more usage of computation and memory, which can be improved by performing OpenMP parallel implementation in our future work.~\cite{2021Hybrid} Therefore, this improved K-means clustering algorithm can accurately and efficiently accelerate large-scale and long-time \textit{ab initio} molecular dynamics with complex-valued hybrid DFT calculations.

\section{Conclusion and outlook} \label{sec:Conclusion}

In conclusion, we present an improved K-means clustering algorithm for complex-valued wavefunctions to select interpolation sampling points in the ISDF decomposition. By applying the new K-means clustering algorithm with SSM into hybrid density functional calculations, we demonstrate that the improved K-means clustering algorithm yields more accurate and smoother interpolation sampling points compared to the K-means clustering with PSM for complex-valued Kohn-Sham orbitals. In particular, K-means with SSM exhibits less loss of accuracy and better stability for AIMD RDF and power spectrum simulations with hybrid density functional. Moreover, we implement the parallel ISDF decomposition for large-scale hybrid functional calculations. We show that the ISDF can scale up to 5,504 CPU cores for a system containing 2,744 atoms. The complex-valued wavefunction is indispensable for multi-k points sampling DFT and RT-TDDFT. The K-means clustering algorithm is more suitable than QRCP method for the dynamic simulation process because of its cheap cost. Therefore, we will apply the complex-valued K-means clustering algorithm to excited-state RT-TDDFT with hybrid functionals in our future work.

%\vspace{3ex}
\section*{Acknowledgments}

This work is partly supported by the Strategic Priority Research Program of the Chinese Academy of Sciences (XDB0450101), the Innovation Program for Quantum Science and Technology (2021ZD0303306), the National Natural Science Foundation of China (22288201, 22173093, 21688102), by the Anhui Provincial Key Research and Development Program (2022a05020052), the National Key Research and Development Program of China (2016YFA0200604, 2021YFB0300600), and the CAS Project for Young Scientists in Basic Research (YSBR-005). The authors thank the Hefei Advanced Computing Center, the Supercomputing Center of Chinese Academy of Sciences, the Supercomputing Center of USTC, the National Supercomputing Center in Wuxi, and Tianjin, Shanghai, and Guangzhou Supercomputing Centers for the computational resources.

\section*{Data Availability}
The data that support the findings of this study are
available from the corresponding author upon reasonable
request.

\appendix

\section{Verification of the feasibility for K-means with SSM}\label{sec:append}
In order to verify the feasibility of SSM as the weight function, here we demonstrate that the interpolation points using K-means with SSM approximately minimize the  residual for the ISDF decomposition. For simplicity, suppose $N = N_{\phi} = N_{\psi}$, transposed Khatri-Rao product $Z$ is $Z(\mathbf{r}) = [ \phi_i(\mathbf{r})\psi_j^\ast(\mathbf{r})]_{i,j=1}^{N}$. We cluster $N_r$ matrix rows of Z into subsets $\{C_{\mu}\}_{\mu = 1}^{N_{\mu}}$ and select $N_{\mu}$ matrix row $Z(\mathbf{r}_{\mu})$ for representing each $C_{\mu}$. Thus the error of ISDF can be approximated as 
\begin{equation}
R = \sum_{\mu=1}^{N_{\mu}}\sum_{\mathbf{r_k} \in C_{\mu}} ||Z(\mathbf{r_k}) - Proj_{span\{Z(\mathbf{r}_{\mu})\}} Z(\mathbf{r_k})||^2
\label{eq:errorISDF}
\end{equation}
where the projection is the $L^2$ inner product
\begin{equation}
Proj_{span\{Z(\mathbf{r}_{\mu})\}} Z(\mathbf{r_k}) = \frac{Z(\mathbf{r_k}) \cdot Z^\ast(\mathbf{r}_{\mu})}{Z(\mathbf{r}_{\mu}) \cdot Z^\ast(\mathbf{r}_{\mu})} Z(\mathbf{r}_{\mu})
\label{eq:proj}
\end{equation}
we define electron density $\rho(\mathbf{r}_{\mu}) = \sum_{i=1}^N |\phi_i(\mathbf{r}_{\mu})|^2 = \sum_{j=1}^N |\psi_j(\mathbf{r}_{\mu})|^2$, $\Phi(\mathbf{r}) = [\phi_i(\mathbf{r})]_{i=1}^N$, $\Psi(\mathbf{r}) = [\psi_j(\mathbf{r})]_{j=1}^N$.
\begin{equation}
\begin{split}
Z(\mathbf{r}_{\mu}) \cdot Z^\ast(\mathbf{r}_{\mu}) 
&= (\Phi(\mathbf{r}_{\mu}) \cdot \Psi^\ast(\mathbf{r}_{\mu}) )(\Phi^\ast(\mathbf{r}_{\mu}) \cdot \Psi(\mathbf{r}_{\mu}))\\
&= \sum_{i,j=1}^N |\phi_i(\mathbf{r}_{\mu})|^2|\psi_j(\mathbf{r}_{\mu})|^2\\
&= (\sum_{i=1}^N |\phi_i(\mathbf{r}_{\mu})|^2)(\sum_{j=1}^N|\psi_j(\mathbf{r}_{\mu})|^2)\\
&= \rho^2(\mathbf{r}_{\mu})
\end{split}
\label{eq:Zmu}
\end{equation}
we have 
\begin{equation}
\begin{split}
R & = \sum_{\mu=1}^{N_{\mu}}\sum_{\mathbf{r} \in C_{\mu}} ||Z(\mathbf{r_k}) - \frac{Z(\mathbf{r_k}) \cdot Z^\ast(\mathbf{r}_{\mu})}{Z(\mathbf{r}_{\mu}) \cdot Z^\ast(\mathbf{r}_{\mu})} Z(\mathbf{r}_{\mu})||^2 \\
& = \sum_{\mu=1}^{N_{\mu}}\sum_{\mathbf{r} \in C_{\mu}} Z(\mathbf{r_k}) \cdot Z^\ast(\mathbf{r_k})[ 1 - \frac{(Z^\ast(\mathbf{r_k}) \cdot Z(\mathbf{r}_{\mu}))(Z(\mathbf{r_k}) \cdot Z^\ast(\mathbf{r}_{\mu}))}{(Z(\mathbf{r_k}) \cdot Z^\ast(\mathbf{r_k})) (Z(\mathbf{r}_{\mu}) \cdot Z^\ast(\mathbf{r}_{\mu}))} ] \\
& = \sum_{\mu=1}^{N_{\mu}}\sum_{\mathbf{r} \in C_{\mu}} \rho^2(\mathbf{r}_{k}) [1 - \frac{(\Phi(\mathbf{r}_{k}) \cdot \Phi^\ast(\mathbf{r}_{\mu}) )^2(\Psi(\mathbf{r}_{k}) \cdot \Psi^\ast(\mathbf{r}_{\mu}) )^2}{\rho^2(\mathbf{r}_{k})\rho^2(\mathbf{r}_{\mu})}] \\
& =  \sum_{\mu=1}^{N_{\mu}}\sum_{\mathbf{r} \in C_{\mu}} \rho^2(\mathbf{r}_{k}) [1 - cos^2(\theta_1(\mathbf{r}_{k}, \mathbf{r}_{\mu}))cos^2(\theta_2(\mathbf{r}_{k}, \mathbf{r}_{\mu}))] \\
& =  \sum_{\mu=1}^{N_{\mu}}\sum_{\mathbf{r} \in C_{\mu}} \rho^2(\mathbf{r}_{k}) [sin^2(\theta_1(\mathbf{r}_{k}, \mathbf{r}_{\mu})) + sin^2(\theta_2(\mathbf{r}_{k}, \mathbf{r}_{\mu})) \quad ...\\ & \quad \quad \quad \quad - sin^2(\theta_1(\mathbf{r}_{k}, \mathbf{r}_{\mu}))sin^2(\theta_2(\mathbf{r}_{k}, \mathbf{r}_{\mu}))] \\
& \leq  \sum_{\mu=1}^{N_{\mu}}\sum_{\mathbf{r} \in C_{\mu}} \rho^2(\mathbf{r}_{k}) [sin^2(\theta_1(\mathbf{r}_{k}, \mathbf{r}_{\mu})) + sin^2(\theta_2(\mathbf{r}_{k}, \mathbf{r}_{\mu}))]
%& \approx  \sum_{\mu=1}^{N_{\mu}}\sum_{\mathbf{r} \in C_{\mu}} \rho^2(\mathbf{r}_{k}) (1 - cos^4(\theta(\mathbf{r}_{k}, \mathbf{r}_{\mu})))
\end{split}
\label{eq:error2}
\end{equation}
where $\theta_1(\mathbf{r}_{k}, \mathbf{r}_{\mu})$ and $\theta_2(\mathbf{r}_{k}, \mathbf{r}_{\mu})$ are the angles between the vectors $\Phi(\mathbf{r}_{k})$ and $\Phi^\ast(\mathbf{r}_{\mu})$ as well as $\Psi(\mathbf{r}_{k})$ and $\Psi^\ast(\mathbf{r}_{\mu})$, respectively. Because
%Here we assume $\theta(\mathbf{r}_{k}, \mathbf{r}_{\mu}) \approx \theta_1(\mathbf{r}_{k}, \mathbf{r}_{\mu}) \approx \theta_2(\mathbf{r}_{k}, \mathbf{r}_{\mu})$. 
\begin{equation}
\begin{split}
& \rho(\mathbf{r}_{k}) [sin^2(\theta_1(\mathbf{r}_{k}, \mathbf{r}_{\mu})) + sin^2(\theta_2(\mathbf{r}_{k}, \mathbf{r}_{\mu}))] \\
& = \Phi(\mathbf{r}_{k}) \cdot \Phi^\ast(\mathbf{r}_{k}) sin^2(\theta_1(\mathbf{r}_{k}, \mathbf{r}_{\mu})) + \Psi(\mathbf{r}_{k}) \cdot \Psi^\ast(\mathbf{r}_{k}) sin^2(\theta_2(\mathbf{r}_{k}, \mathbf{r}_{\mu}))\\
& \leq ||\Phi(\mathbf{r}_{k}) - \Phi^\ast(\mathbf{r}_{\mu})||^2 + ||\Psi(\mathbf{r}_{k}) - \Psi^\ast(\mathbf{r}_{\mu})||^2
\end{split}
\label{eq:cos}
\end{equation}
we can obtain
\begin{equation}
\begin{split}
R & \leq  \sum_{\mu=1}^{N_{\mu}}\sum_{\mathbf{r} \in C_{\mu}} \rho(\mathbf{r}_{k})
[||\Phi(\mathbf{r}_{k}) - \Phi^\ast(\mathbf{r}_{\mu})||^2 + ||\Psi(\mathbf{r}_{k}) - \Psi^\ast(\mathbf{r}_{\mu})||^2]\\
& \approx \sum_{\mu=1}^{N_{\mu}}\sum_{\mathbf{r} \in C_{\mu}} \rho(\mathbf{r}_{k}) (||\nabla_r\Phi(\mathbf{r}_{\mu})||^2 + ||\nabla_r\Psi(\mathbf{r}_{\mu})||^2)
||\mathbf{r}_{k} - \mathbf{r}_{\mu}||^2 \\
& =  \frac{1}{2}\sum_{\mu=1}^{N_{\mu}}\sum_{\mathbf{r} \in C_{\mu}} (\sum_{i=1}^N |\phi_i(\mathbf{r}_{k})|^2 + \sum_{j=1}^N|\psi_j(\mathbf{r}_{k})|^2) \quad ...\\
& \quad \quad \quad \quad (||\nabla_r\Phi(\mathbf{r}_{\mu})||^2 + ||\nabla_r\Psi(\mathbf{r}_{\mu})||^2)
||\mathbf{r}_{k} - \mathbf{r}_{\mu}||^2
\end{split}
\label{eq:error3}
\end{equation}
Thus the minimization criterion of weighted K-means with SSM can be derived when the spatial inhomogeneity of the gradient $\Phi(\mathbf{r})$ and $\Psi(\mathbf{r})$ is neglected.

\bibliography{achemso}

\vspace{3ex}
\[
\includegraphics[width=\textwidth]{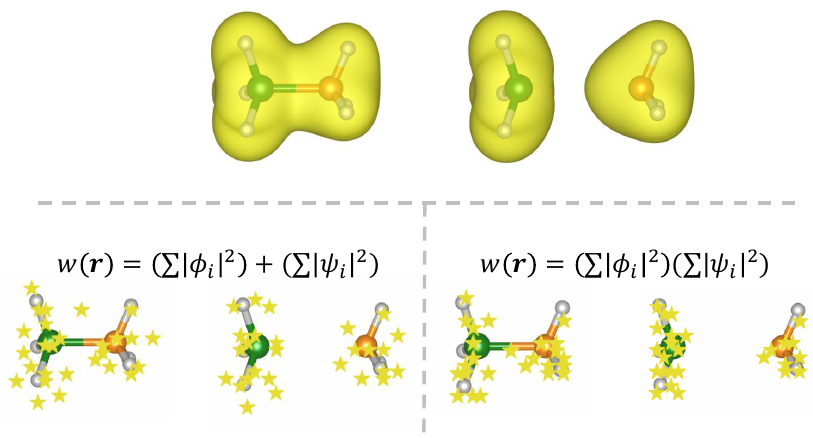}
\]
\centerline{TOC graphic}

\end{document}